\documentclass{emulateapj}
\slugcomment{Accepted to ApJ; Jun 29 2014}

\pdfoutput=1
\newcommand{\Gcloud}{G34.43$+$00.24}
\newcommand{\Msun}{\ensuremath{M_{\odot}}}
\newcommand{\mmthree}{Field 6}
\newcommand{\fonethreeone}{Field 1}

\usepackage[backref,breaklinks,colorlinks,citecolor=blue]{hyperref}
\usepackage[all]{hypcap} 

\shorttitle{Low-mass Protostars in G34.43}
\shortauthors{Foster, J.B. et al.}

\begin{document}

\title{Distributed Low-Mass Star Formation in the IRDC \Gcloud}

\author{
Jonathan B. Foster\altaffilmark{1}, 
H\'{e}ctor G. Arce\altaffilmark{2}, 
Marc Kassis\altaffilmark{3},
Patricio Sanhueza\altaffilmark{4},
James M. Jackson\altaffilmark{4},
Susanna C. Finn\altaffilmark{5},
Stella Offner\altaffilmark{2},
Takeshi Sakai\altaffilmark{6},
Nami Sakai\altaffilmark{7},
Satoshi Yamamoto\altaffilmark{7},
Andr\'{e}s E. Guzm\'{a}n\altaffilmark{8},
Jill M. Rathborne\altaffilmark{9}
}

\altaffiltext{1}{Yale Center for Astronomy and Astrophysics, New Haven, CT 06520, USA; \href{mailto:jonathan.b.foster@yale.edu}{jonathan.b.foster@yale.edu}}
\altaffiltext{2}{Department of Astronomy, Yale University, P.O. Box 208101, New Haven, CT 06520, USA}
\altaffiltext{3}{W. M. Keck Observatory, 65-1120 Mamalahoa Highway, Kamuela, HI 96743, USA}
\altaffiltext{4}{Institute for Astrophysical Research, Boston University, Boston, MA 02215, USA}
\altaffiltext{5}{Center for Atmospheric Research, UMASS Lowell, 600 Suffolk St., Lowell MA 01854, USA}
\altaffiltext{6}{Graduate School of Informatics and Engineering, The University of Electro-Communications, Chofu, Tokyo 182-8585, Japan}
\altaffiltext{7}{Department of Physics, The University of Tokyo, Bunkyo-ku, Tokyo 113-0033, Japan}
\altaffiltext{8}{Harvard-Smithsonian Center for Astrophysics, 60 Garden Street, Cambridge, MA 02138, USA}
\altaffiltext{9}{CSIRO Astronomy and Space Science, P.O. Box 76, Epping NSW 1710, Australia}

\begin{abstract}
We have used deep near-infrared observations with adaptive optics to discover a distributed population of low-mass protostars within the filamentary Infrared Dark Cloud \Gcloud. We use maps of dust emission at multiple wavelengths to determine the column density structure of the cloud. In combination with an empirically verified model of the magnitude distribution of background stars, this column density map allows us to reliably determine overdensities of red sources that are due to embedded protostars in the cloud. We also identify protostars through their extended emission in the $K$ band which comes from excited H$_2$ in protostellar outflows or reflection nebulosity. We find a population of distributed low-mass protostars, suggesting that low-mass protostars may form earlier than, or contemporaneously with, high-mass protostars in such a filament. The low-mass protostellar population may also produce the narrow line-width SiO emission observed in some clouds without high-mass protostars. Finally, we use a molecular line map of the cloud to determine the virial parameter per unit length along the filament and find that the highest mass protostars form in the most bound portion of the filament, as suggested by theoretical models. 

\end{abstract}

\keywords{}

\section{Introduction}
\label{sec:intro}
The initial mass function (IMF) of stars is, within observational errors, invariant locally \citep{Offner:2013}; there is an ongoing debate about whether or not a constant IMF reproduces observations of high-redshift galaxies \citep{Bastian:2010} and the mass function of faint dwarf galaxies \citep{Geha:2013}. The steep slope of the IMF means that many low-mass stars form in a cluster with high-mass stars (i.e., stars with masses greater than 8 \Msun), a fact that has strong implications for the origin of planetary systems \citep{Looney:2006}. It is possible that low-mass regions such as Taurus have a different shape to the IMF, with more low-mass stars \citep{Luhman:2009}. Although there may be many such low-mass regions, each forms only a small number of stars; the IMF is strongly influenced by regions of high-mass star formation.

The observations of \emph{Herschel} have revived interest in the study of filaments as one of the primary modes into which dense matter organizes itself under gravity and turbulence \citep{Molinari:2010}. These large filaments can have aspect ratios of nearly 100:1, with high-mass star formation occurring in the beads (or clumps) along the filamentary string \citep{Jackson:2010}. Such filaments, which can be traced in mid-infrared extinction, appear to be present in virtually all regions of high-mass star formation, including the famous integral-shaped filament in Orion \citep{Johnstone:1999}. High-mass stars form, predominantly, in filaments.

It follows that the IMF is set by star formation in filaments forming high-mass stars -- but questions remain. (1) Do low- and high-mass stars form coevally or not? (2) Is the IMF universal within a filament, or is it different in the dense clumps and more diffuse inter-clump regions?

Despite numerous years of study, the temporal sequence of star formation in molecular clouds remains highly uncertain. The large uncertainty associated with the ages of young stars \citep[e.g., see the recent revision to ages in][]{Bell:2013} is a significant factor. For high-mass star formation, the question is particularly unsettled since it is very hard to estimate the age of high-mass protostars. Since high-mass stars are almost always seen in the presence of a cluster of low-mass stars \citep[and the presence of truly isolated high-mass stars, e.g.,][ is debated]{Oey:2013}, it is natural to ask whether the high-mass stars form before, after, or coevally with the low-mass stars. By searching for low-mass protostars in the earliest phases of high-mass star formation one may gain insight into the temporal sequence without the need to determine accurate stellar ages.

There could be two modes of star formation present within a filament: clustered star formation in the dense clumps and Taurus-like distributed star formation in the inter-clump medium. In simulations of turbulent filamentary clouds, \citet{Bonnell:2011} distinguish between dense, bound portions of the filament and unbound portions. The star formation efficiency varies dramatically between these two regimes, ranging from 40\% in the bound regions to less than 1\% in the unbound regions. In addition, the IMF is different; it is strongly peaked around the local Jeans mass in the unbound region, while the IMF is normal \citep[i.e. consistent with a log-normal distribution at low masses and a power-law above 1 \Msun\ as in][]{Chabrier:2003} in the bound portions.

The fundamental question is, on what temporal and spatial scale is the IMF set within a star-forming region? The answer will constrain our theoretical understanding of star formation and the shape of the IMF in conditions quite different from those found locally in our Galactic neighborhood. These questions can be answered by observing the low-mass protostellar population in the earliest stages of high-mass star formation. 

\begin{figure}
\includegraphics[width=0.48\textwidth]{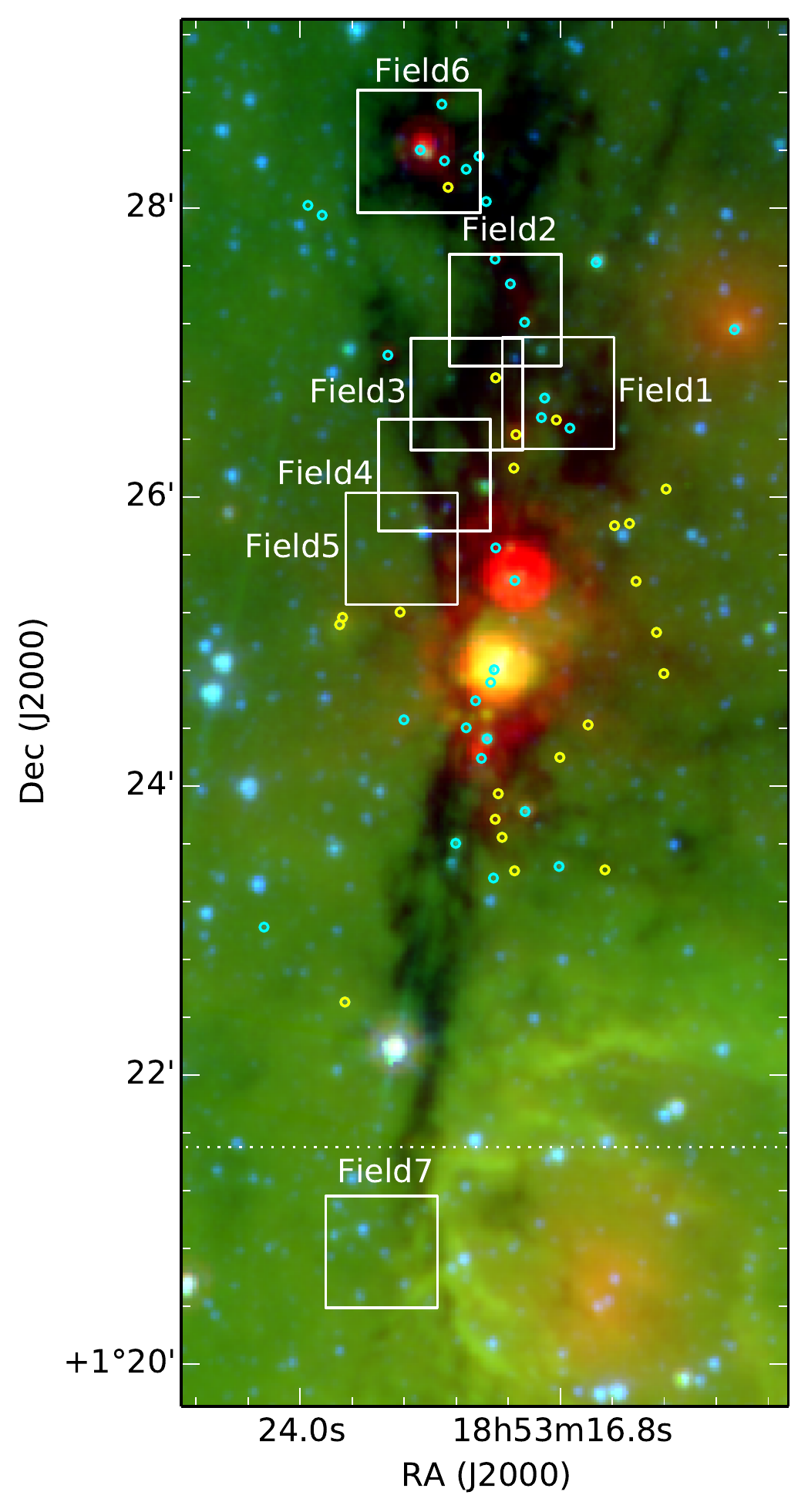}
\caption{Overview of the cloud \Gcloud\ on a GLIMPSE and MIPSGAL three-color image with 4.5 (blue), 8.0 (green) and 24 \micron\ (red). White rectangles show the fields observed in the $K$ band and $H$ band in this work. Candidate YSOs from \citet{Shepherd:2007} are shown in cyan (sources with a spectral energy distribution fit) and yellow (24 \micron\ point sources without a full fit); that study considered \emph{Spitzer} sources everywhere in this image except south of the white dotted line.}
\label{fig:spitzer_overview}
\end{figure}

\Gcloud\ is a well-studied filamentary Infrared Dark Cloud (IRDC), 9\arcmin\ in extent and containing young high-mass stars and ultra-compact H \textsc{ii} regions \citep{Miralles:1994, Shepherd:2004, Rathborne:2005}. \autoref{fig:spitzer_overview} provides an overview of \Gcloud. For thorough reviews of the previous studies on this cloud see \citet{Shepherd:2007} and \citet{Sanhueza:2010}. At a distance of 3.7 kpc, the cloud contains roughly 1000 \Msun\ of molecular gas \citep{Miralles:1994}.

The distance to \Gcloud\ is a matter of some debate. Parallax measurements of H$_2$O maser sources within the cloud from VERA \citep[VLBI Exploration of Radio Astrometry, ][]{Kurayama:2011} place the cloud at 1.56$^{+0.12}_{-0.11}$ kpc, roughly half the 3.7-3.9 kpc value estimated from kinematic distances \citep[e.g.,][]{Rathborne:2005, Zhang:2005, Sanhueza:2012} and from near-infrared extinction \citep{Foster:2011}. 

Although a parallax measurement would seem to be the more reliable distance determination, it is important to note that parallax determinations to the same source do not always give good agreement. \citet{Nagayama:2011} and \citet{Zhang:2013} determine inconsistent distances to G048.61+0.02, 5.03$\pm$0.19 kpc versus 10.75$^{+0.61}_{-0.55}$ kpc respectively. In that case the \citet{Zhang:2013} measurement is in agreement with the kinematic distance and they suggest that the \citet{Nagayama:2011} measurement is wrong because both background reference sources used in that work are offset a similar distance in the same direction with respect to the source, rendering their analysis vulnerable to systematic errors due to unmodeled tropospheric delays. The \citet{Kurayama:2011} measurement relied on only a single reference background source, and the measurements are inherently difficult due to the low declination of this target. For this reason, and because of the challenge in understanding the radial velocity of \Gcloud\ if it were at 1.56 kpc, we adopt the kinematic/extinction distance of 3.9 kpc for this cloud. Note that \citet{Sakai:2013} use a distance of 1.56 kpc.

\citet{Shepherd:2007} have conducted the best previous search for protostars in \Gcloud, using $Spitzer$ GLIMPSE \citep{Benjamin:2003} and MIPSGAL \citep{Carey:2009} data to fit for protostellar properties using the \citet{Robitaille:2006} models. In addition, \cite{Shepherd:2004} detected a low-mass population of stars around one of the ultra-compact H \textsc{ii} regions within the cloud. On the basis of the mass of the sources in the ultra-compact H \textsc{ii} regions, \citet{Shepherd:2007} suggest that these high-mass stars are 10$^5$ yr old, while the low-mass protostars are 0.3 - 3 Myr old -- potentially older, but all these ages are very uncertain. 

In this work, we demonstrate that deep near-infrared observations (particularly with adaptive optics) can reveal the low-mass protostellar distribution in IRDCs. We observe \Gcloud\ and demonstrate the presence of distributed, low-mass star formation occurring between the dense millimeter clumps that have sufficient mass to form a star cluster with a high-mass star. In addition, the presence of low-mass protostars near one of the dense clumps that does not yet have signatures of a high-mass protostar suggests that, on the scale of a few parsecs, low-mass stars may form before or contemporaneously with high-mass stars. In combination with a molecular line map and dust maps of the filament, we show that high-mass stars predominantly form in the most bound parts of the filaments, while we find the low-mass protostars in less bound portions of the filament. 

\section{Observations}

\subsection{$H$ and $K$ Imaging with Keck}
\capstartfalse
\begin{deluxetable}{lcccc}
\tablewidth{0pc}
\tablecaption{Fields Observed with Keck}
\tablehead{
\colhead{Name} & \colhead{R.A.} & \colhead{Decl.} & \colhead{K$_{\mathrm{lim}}$} & \colhead{H$_{\mathrm{lim}}$}\\
	& hh:mm:ss & dd:mm:ss & &
}
\startdata

Field 1 & 18:53:16.87 & 1:26:43.08 & 21.5 & 24.4 \\
Field 2 & 18:53:18.34 & 1:27:17.64 & 21.5 & 24.5 \\
Field 3 & 18:53:19.39 & 1:26:42.72 & 21.6 & 24.5 \\
Field 4 & 18:53:19.39 & 1:26:09.24 & 21.1 & 23.9\\
Field 5 & 18:53:21.17 & 1:25:38.64 & 21.2 & 24.0\\
Field 6 & 18:53:20.74 & 1:28:23.88 & 21.8 & 25.1\\
Field 7 & 18:53:21.74 & 1:20:46.68 & 20.9 & 23.8\\

\enddata
\label{table:fields}
\tablecomments{Completeness limits are estimated to the nearest 0.1 mag by injection and recovery of artificial stars near the center of the field. The quoted limits are for 99\% recovery of injected stars.}
\end{deluxetable}
\capstarttrue

\begin{figure}
\includegraphics[width=0.48\textwidth]{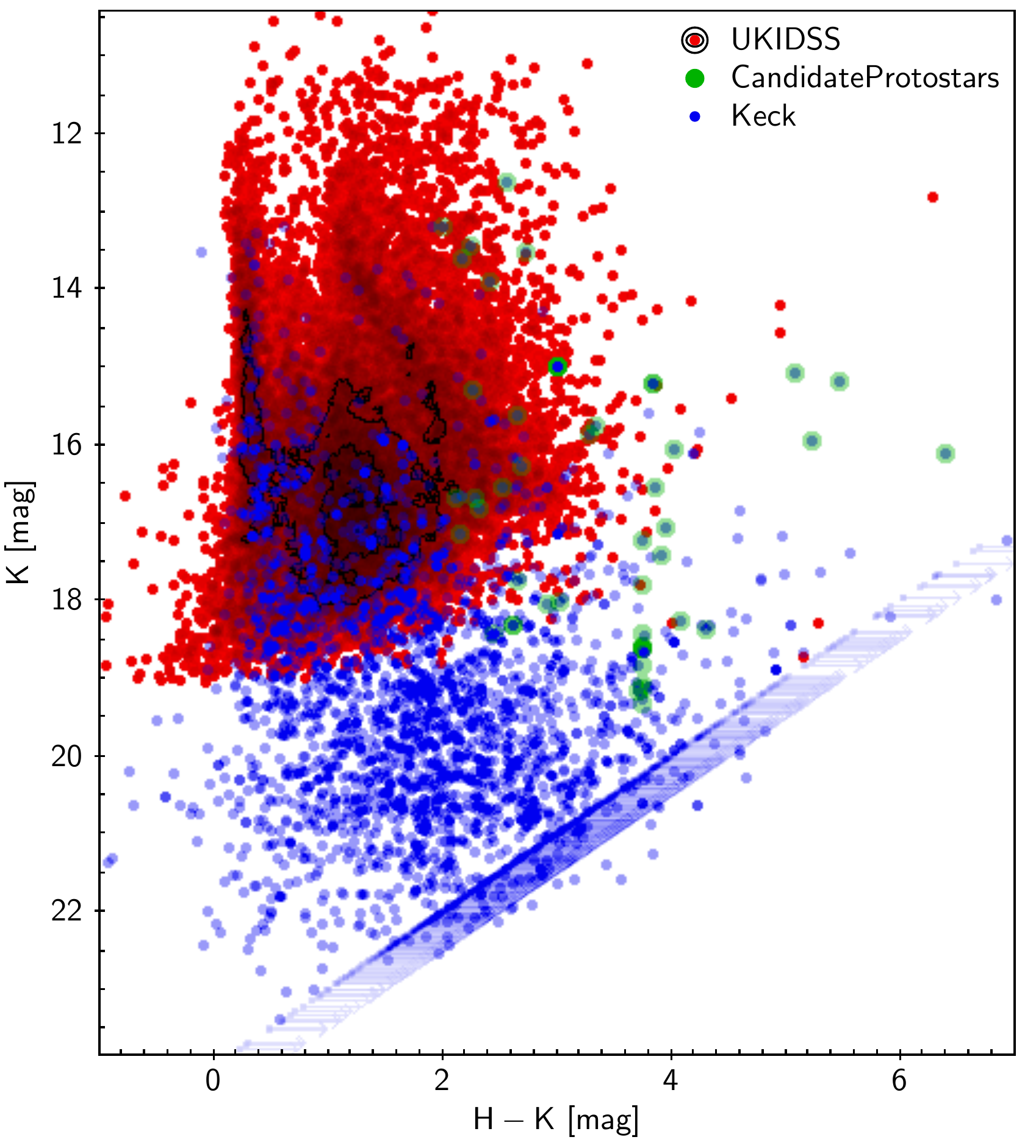}
\caption{Color-magnitude diagram of both UKIDSS sources (red; within 0\arcdeg\.1 radius of \Gcloud) and Keck sources (blue; as sampled on the filament in the fields shown in \autoref{fig:spitzer_overview}). Keck sources with no $H$ band detection are displayed as limits, and UKIDSS sources without $H$ band detections are suppressed for clarity. Highlighted in green are the protostar candidates identified as described in \autoref{sec:identification}.}
\label{fig:cmd}
\end{figure}

The near-infrared images were taken using NIRC2 on Keck II with adaptive optics.  We took all observations using the wide camera (40 mas pixel$^{-1}$ and 40$\times$40\arcsec\ field of view). Field names and locations are shown in \autoref{fig:spitzer_overview} and \autoref{table:fields}. We observed Field 6 on August 20, 2010 using laser guide star adaptive optics and imaging in $K_p$ and $H$ filters. We took 3 s exposures with 30 coadds in $K_p$ and 3 s exposures with 10 coadds in $H$. The total exposure was 1 hr in $K_p$ and 14 minutes in $H$. Although $K_p$ is slightly bluer and wider than the standard $K$ filter, this distinction is unimportant in this analysis, and henceforth we shall refer to these as $K$-band observations.

We observed six other fields on June 21, 2013. The positions of these fields were dictated by the presence of stars that could be used as natural guide stars for adaptive optics. The guide star requirement of $R$ band $<$ 14 mag effectively limited us to bright, blue foreground stars; as such, the fields are relatively random samples of the cloud. We observed in both the $K$ band and $H$ band, using 15 s exposures with six coadds at each position. Total exposure time was not the same for every field, and ranged between 15 and 22.5 minutes in $K$ and between 15 and 30 minutes in $H$. We increased our observing efficiency by observing multiple fields around each natural guide star and interleaving the observations of different fields so that we did not have to re-acquire the guide star for each field. At each position we did a five-point dither pattern with 5\arcsec\ offsets. 

The data were reduced in a standard fashion by subtracting dark frames, dividing by a flat, and then performing sky subtraction using a running median of five individual exposures. We also applied linearity corrections, a distortion solution, and a cosmic-ray removal algorithm. During both nights of observations, the $K$-band images showed variable excess background in the lower-right portion of the image. To remove this excess background, we subtracted a smoothed version (using a 200 pixel smoothing kernel) of each image from itself prior to final coaddition. This corresponds to an 8\arcsec\ smoothing kernel, and therefore any structures larger than this are suppressed (although no such structures appear to be present when we reduce the images without subtracting off a smoothed image). 

Aperture photometry was performed using Source Extractor \citep{Bertin:1996}; an aperture of 1\arcsec\ was used. World Coordinate System (WCS) solutions and photometric calibration were based on the UKIRT Infrared Deep Sky Survey \citep[UKIDSS;][]{Lawrence:2007} Galactic Plane Survey \citep[GPS;][]{Lucas:2008}.  Since we attempted to optimize the surface area of the cloud accessible with a limited number of guide stars, our guide stars tended to be near the edges of our images. As such, the images show some deformation in the sense that the point-spread function of stars near the edge of the image are stretched toward the guide star. We fit extra WCS components as a sixth-order polynomial (using the Simple Imaging Polynomial (SIP) convention) to account for this deformation. In some cases there were not enough UKIDSS stars to well-constrain the fit for these additional components, and in this case there is often a small ($<$0\arcsec.5) offset between the $K$ and $H$-band images in some portion of the image. The resulting color-magnitude diagram is shown in \autoref{fig:cmd}.

Completeness limits for the fields are estimated by injection and recovery of artificial stars and are reported in \autoref{table:fields}. We inject stars with 0.1 mag steps and quote the magnitude at which we recover 99\% of the injected stars. These completeness limits do not account for the variability in point-spread function as a function of distance from the guide star within each image, which appears to have only a small effect on the completeness limit. In addition, the edges of the images have some additional noise, both from reduced coverage (due to the dither) and due to the variable $K$-band excess background in the lower-right portion of the detector.

\begin{figure}
\includegraphics[width=0.48\textwidth]{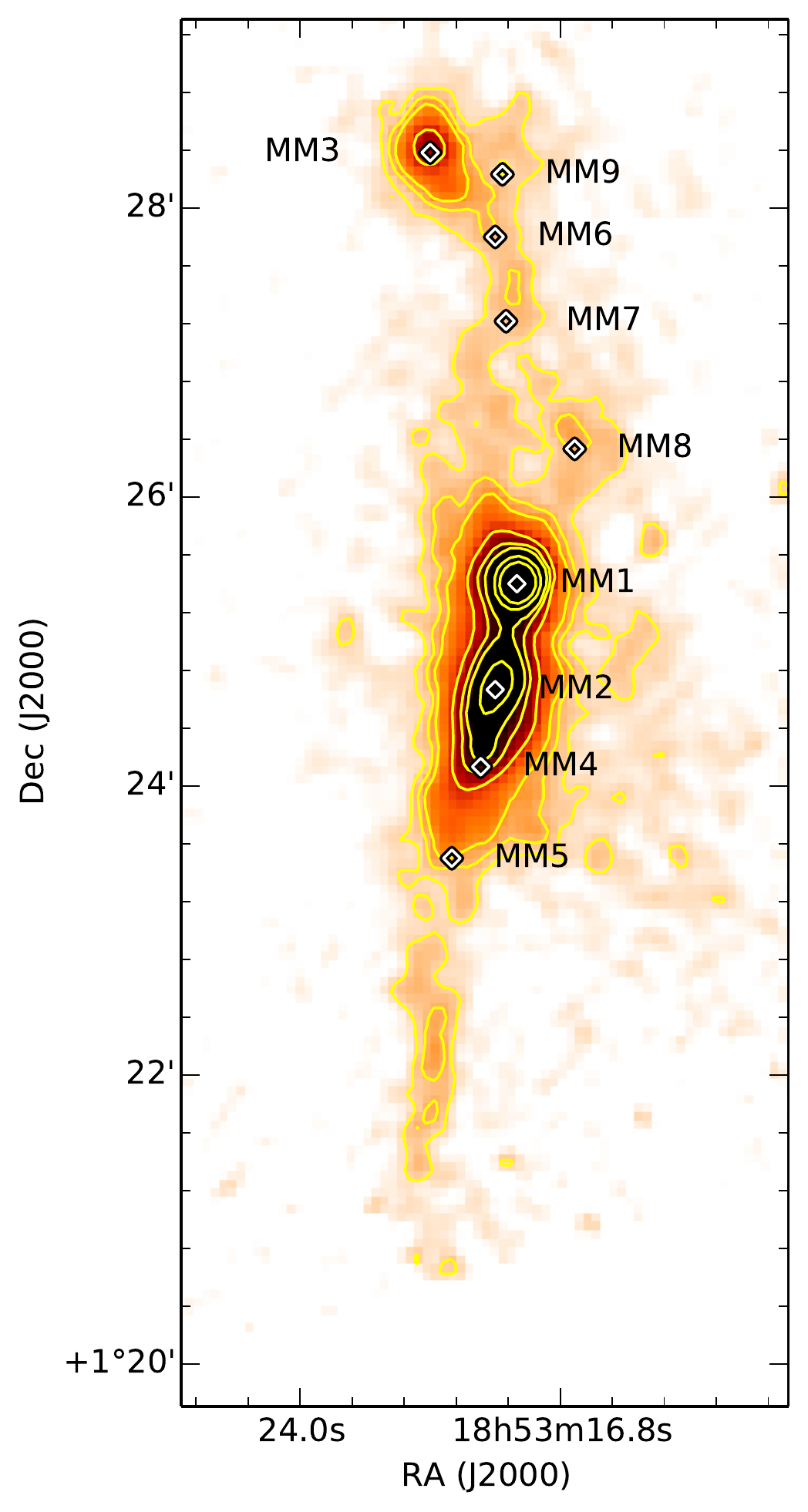}
\caption{MAMBO 1.2 mm map of \Gcloud\ with contours at 60, 90, 120, 240, 360, 480, 840, and 1200 mJy beam$^{-1}$. The millimeter cores identified by \citet{Rathborne:2006} are shown as white diamonds and are labeled as therein (MM1 to MM9).}
\label{fig:mambo_overview}
\end{figure}

\subsection{UKIDSS GPS}

Since our Keck imaging does not cover the full cloud, we also make use of UKIDSS GPS data to search for protostars in the portion of the cloud not covered by our Keck observations and to empirically estimate the background and foreground luminosity function for sources toward the cloud. The UKIDSS project is defined in \citet{Lawrence:2007}. UKIDSS uses the UKIRT Wide Field Camera \citep[WFCAM; ][]{Casali:2007}. The photometric system is described in \citet{Hewett:2006}, and the calibration is described in \citet{Hodgkin:2009}. The pipeline processing and science archive are described in \citet{Hambly:2008}. We used Data Release 8 downloaded from the WFCAM Science Archive.

\subsection{$K$ band Spectra with IRTF}

As part of a larger survey, we obtained near-infrared spectra of the brightest extended $K$ band sources in Field 6 using SPeX on the IRTF. Observations were taken on June 14-15, 2011 and June 15, 2012. Weather during the 2011 run was partly cloudy with good seeing (0\arcsec.7 - 0\arcsec.4). Weather during the 2012 run was cloudy with 1\arcsec\ seeing. Sources were observed with 120 s exposures and either three or six cycles nodding between two positions in the slit. We observed in 0.8-2.5~\micron\ cross-dispersed mode with a 0\arcsec\.5$\times$15\arcsec\ slit with an R of 1200 and interspersed observations with observations of A0V stars for telluric correction.

Reduction was performed using Spextool \citep{Cushing:2004,Vacca:2003} and followed normal reduction procedures. We used extended source extraction with 2\arcsec\ extraction windows. The weak or non-existent continuum level for some sources in the NIR means that division by the telluric standard increased the noise (since these objects had little continuum flux, there was insignificant telluric absorption because there was nothing to absorb). The $K$ band portion of our spectra has relatively weak telluric lines, and there is rarely any detected emission at shorter wavelengths. Since we are concerned primarily with line emission, we simply blank out the portions of the spectra where telluric correction is significant; this allows us to focus on the real line emission.

\subsection{NH$_3$ with the GBT}
We obtained NH$_3$ data for \Gcloud\ using the new $K$-band Focal Plane Array (KFPA) on the Green Bank Telescope (GBT).\footnote{N.B. This is the radio $K$ band, not the NIR $K$ band referenced widely throughout the rest of this work.} The KFPA is a 7-beam array of receivers which provides dramatically improved mapping speed. This cloud was mapped in January, 2011 with the para-NH$_3$ (1,1) and (2,2) inversion transitions \citep[at 23.69450 and 23.72263 GHz, respectively;][]{Lovas:2004} observed simultaneously. The GBT has a beam size of 31\arcsec\ at this frequency and we observed with 21.36 KHz resolution, corresponding to 0.275 km s$^{-1}$ velocity resolution at this frequency. We used a main beam efficiency of $\eta_{\mathrm{mb}}$ = 0.81 to place our observations on the $T_{\mathrm{mb}}$ scale. On this scale, the noise is 0.07 K channel$^{-1}$.

\subsection{Continuum Submillimeter/Millimeter Data}

We use the publicly available $Herschel$ Hi-Gal data \citep{Molinari:2010} and the MAMBO IRAM 30 m dust continuum map at 1.2~mm from \citet{Rathborne:2006} in order to derive the column density of the cloud. The MAMBO map has a beam size of 11\arcsec\ and an rms noise level of $\sim$10 mJy beam$^{-1}$ and is shown in \autoref{fig:mambo_overview}. The $Herschel$ Hi-Gal data was produced by combining both cross-scanning directions and re-reducing using the standard procedures in the Herschel Interactive Processing Environment (HIPE v9.2). This data set includes maps at 70, 160, 250, 350 and 500~\micron\ with beam sizes ranging from 9\arcsec\.2 to 37\arcsec \citep{Olmi:2013}. The point source sensitivity of these maps is typically 10-15 mJy \citep{Molinari:2010}.

For one of the regions observed, Field 6, we compare with the ALMA observations presented in \citet{Sakai:2013}, which allow us to obtain a more complete picture of the protostellar content of this clump. These observations were carried out during ALMA Cycle 0 using ALMA Band 6. We use the continuum data from these observations made by averaging the line-free channels of the four 234 MHz basebands between 228 and 245 GHz, which gives the continuum at 1.3 mm. The synthesized beam is 0\arcsec.82$\times$0\arcsec.61 and the rms noise in the continuum image is 0.3-0.4 mJy~beam$^{-1}$.

\section{Estimating Properties of the Filament}

Our basic analysis of the near-infrared data involves determining whether a given source, which is too red to be a foreground star, is likely to be a background star reddened by the dust in the filament. If a red source is unlikely to be a background star it is likely to be an embedded protostar. This assignment requires a column density map of the cloud; this requires knowledge of the temperature of the dust in order to convert submillimeter/millimeter fluxes to column densities. 

The available data sets for estimating the temperature are the $Herschel$ Hi-Gal data and the GBT NH$_3$ data, which are both relatively coarse resolution $> 30$\arcsec\ compared with the MAMBO 1.2 mm map. In order to assess the likelihood that an NIR source is a protostar it is critical to use the highest resolution map possible, as unresolved density structure will introduce errors in the likelihood calculation; unresolved low column density features or holes in the cloud will allow background stars to shine through a region that looks, on average, to be too dense to allow a background star to be visible. 

For this reason, we make the assumption that the dust temperature is relatively slowly varying, fit for the dust temperature at coarse angular resolution, and use these dust temperatures to convert the higher resolution MAMBO 1.2-mm map into a column density map. We outline the steps to achieve this in the following sections.

\subsection{Dust Temperature from $Herschel$}
\label{sec:dustmap}
\begin{figure}
\includegraphics[width=0.48\textwidth]{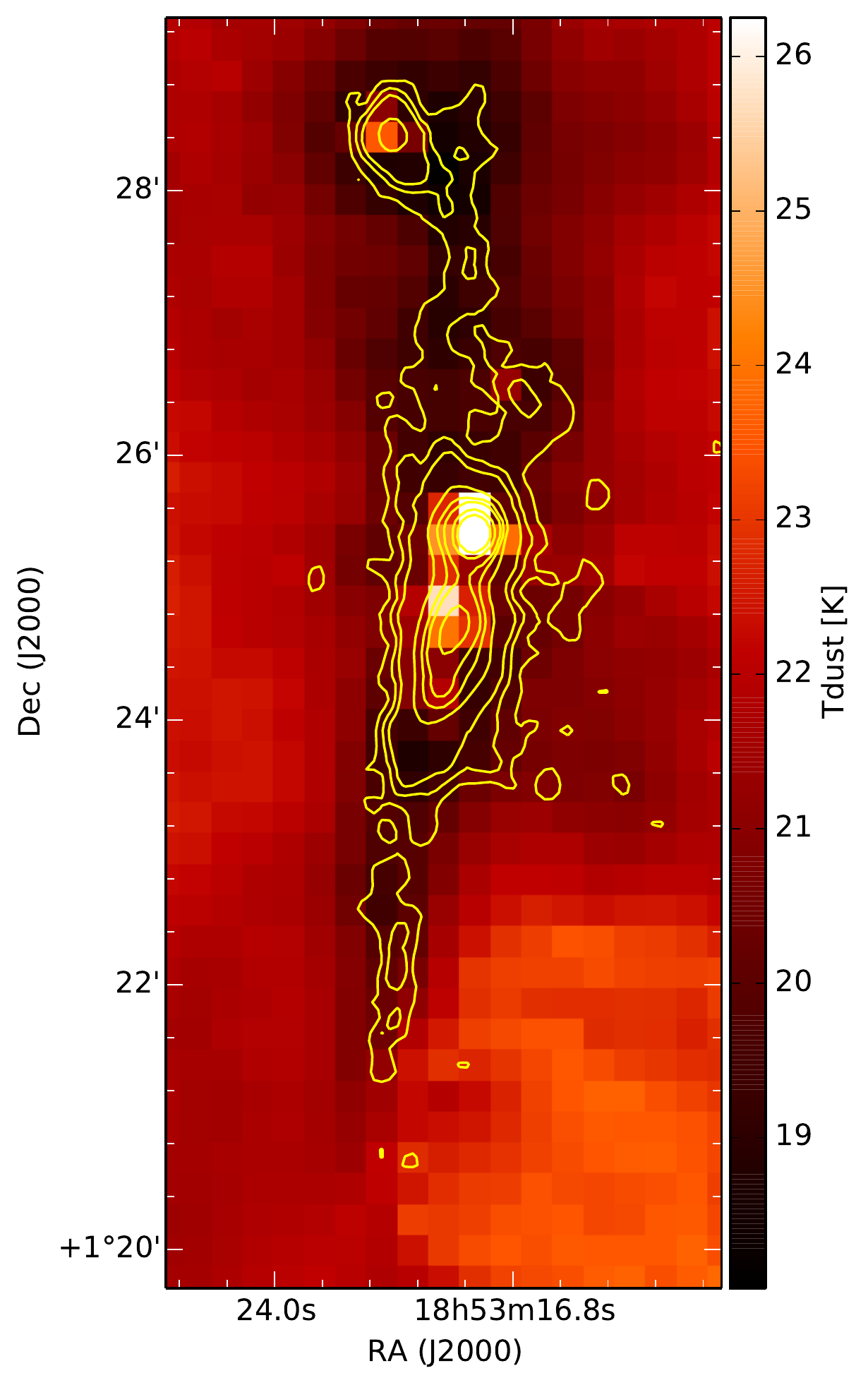}
\caption{Dust temperature estimated from $Herschel$ data. Dust temperatures are calculated (and shown) at the resolution of the coarsest $Herschel$ map (500 \micron) and are then interpolated to higher resolution under the assumption that the dust temperature is slowly varying. The same 1.2~mm contours as in \autoref{fig:mambo_overview} are shown for reference.}
\label{fig:dust_temp}
\end{figure}

We convolve all the $Herschel$ maps and the MAMBO 1.2~mm map to the resolution of the coarsest map (500~\micron, 37\arcsec). We subtract off the median value of each map after masking out \Gcloud, in order to roughly account for the uncertain zero point of the $Herschel$ maps and the background/foreground emission from the rest of the Galactic plane. 

For each pixel, we fit for the dust temperature using a graybody fit, where the opacity/emissivity at each wavelength is taken from the \citet{Ossenkopf:1994} model for dust with thin ice mantles coagulating for 10$^5$ yr at a density of 10$^{6}$ cm$^{-3}$ and assuming a gas-to-dust ratio of 100:1. This differs from the standard procedure of fitting submillimeter emission \citep[e.g.,][]{Hildebrand:1983} because rather than assuming that emission from dust is a blackbody modified by some power law with exponent $\beta$, it allows the dust emissivity to have the more complicated dependence on wavelength estimated by the \citet{Ossenkopf:1994} model. In addition, we do not assume that the emission is optically thin. We fit each flux as

\begin{equation}
\label{eqn:bb}
S_{\nu}^{\mathrm{beam}} = [1- e^{-\tau}]\times[\Omega_{A} B_{\nu} (T)],
\end{equation}
where 
\begin{equation}
\tau = \mu_{H_2} m_H \kappa_{\nu} N,
\end{equation}
rather than assuming $\tau \ll 1$. It is important to include the full optical depth term, as the cloud is not optically thin at 70 and 160~\micron. Here $S_{\nu}^{\mathrm{beam}}$ is the flux per beam, $\Omega_{A}$ is the beam solid angle, $\mu_{\mathrm{H_2}}$ is the mass per hydrogen molecule, $\kappa_{\nu}$ is the dust opacity, and $B_{\nu} (T)$ is the Planck function. We used $\mu_{\mathrm{H_2}} = 2.8$ and $\kappa_{\nu}$ as explained above. The resulting dust temperature map is shown in \autoref{fig:dust_temp}.

We still assume a single dust temperature along the line of sight and that the dust optical properties are constant along the line of sight (and well described by the \cite{Ossenkopf:1994} model), with all the concurrent uncertainty \citep[e.g.,][]{Shetty:2009b}. In particular, toward the high-luminosity embedded sources in \Gcloud\ (MM1, MM2, MM3), the use of a single dust temperature along the line of sight will tend to underestimate the column density of the cloud as this model cannot fit both the warm dust (dominant at 70 and 160~$\micron$ ) and the cold dust (dominant at 1.2~mm); see \citet{Malinen:2011}. The dust temperature correction should be thought of as a first-order correction to the column density calculation. Underestimating the column density only makes one less sensitive to detecting embedded sources (in \autoref{sec:nir}); it does not lead to the false inference that background stars must actually be embedded sources.

\subsection{Line Width from NH$_3$}

\begin{figure}
\includegraphics[width=0.48\textwidth]{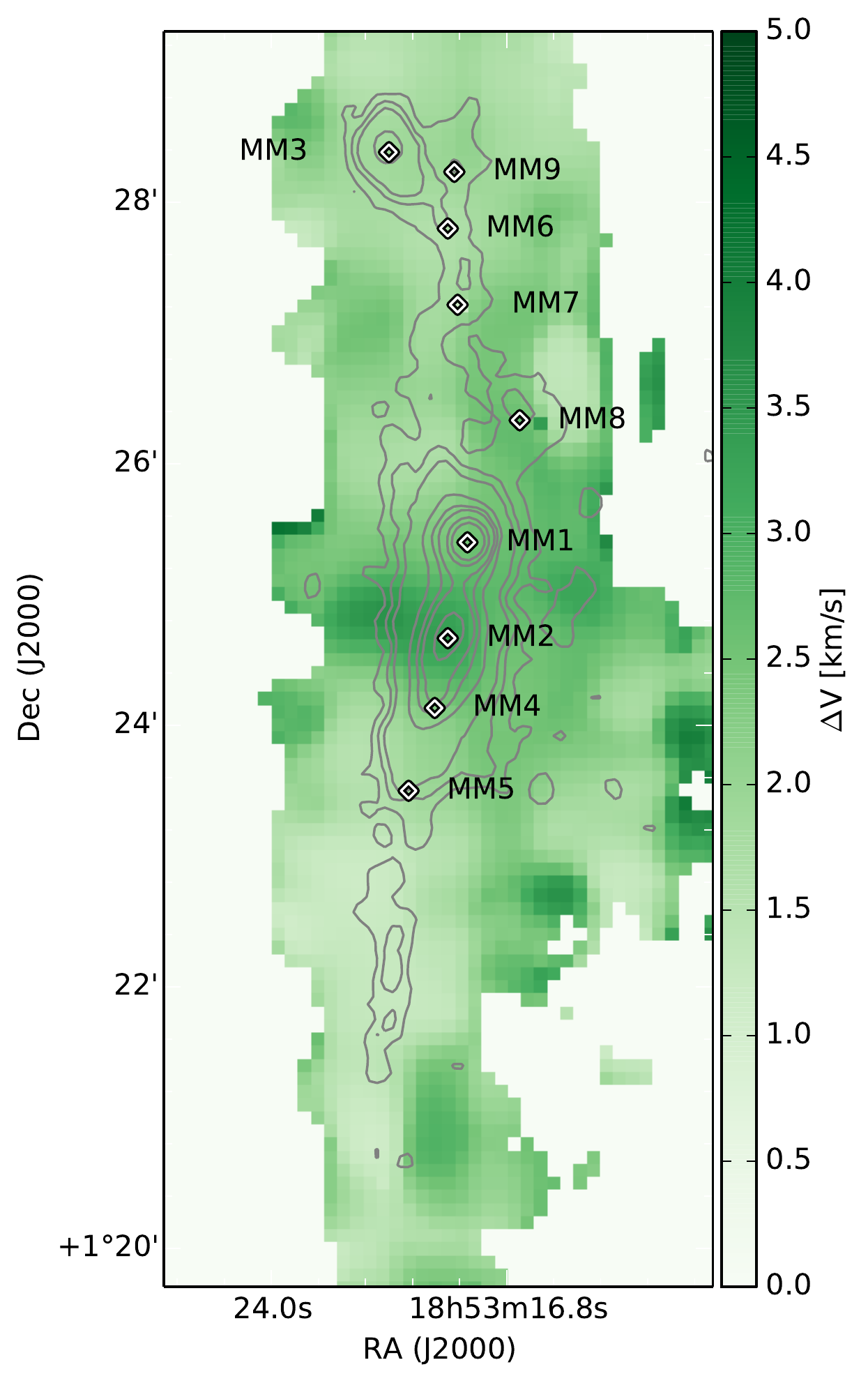}
\caption{FWHM of the NH$_3$ (1,1) transition. The same 1.2~mm contours and clumps as in \autoref{fig:mambo_overview} are shown for reference.}
\label{fig:linewidth}
\end{figure}

We use the NH$_3$ data to calculate the total velocity dispersion of the gas. We forward model the (1,1) spectrum under the assumption of a homogeneous slab with uniform intrinsic velocity dispersion and excitation conditions for all hyperfine transitions \citep{Rosolowsky:2008}. This allows us to correct for the the line broadening resulting from optical depth effects and produces the map of the line width shown in \autoref{fig:linewidth}.

In theory, one can use observations of the NH$_3$ ($J$=1,$K$=1) and ($J$=2, $K$=2) inversion transitions to estimate the kinetic temperature of the gas \citep{Ho:1979, Ho:1983, Walmsley:1983, Maret:2009}. However, the kinetic temperature estimated from the (1,1) and (2,2) transitions underestimates the gas temperature above about 15 K and becomes a progressively worse estimate of the kinetic temperatures at higher (true) kinetic temperatures \citep{Walmsley:1983}. Therefore, we use the dust temperature map calculated in \autoref{sec:dustmap}, rather than a kinetic gas temperature calculated from NH$_3$.

One can use the kinetic temperature of NH$_3$ to correct the observed velocity dispersion for the difference between the thermal line width coming from the heavier NH$_3$ and the lighter H$_2$ \citep[e.g.,][]{Foster:2009} but this correction is small (0.1 km s$^{-1}$ for $T_{\mathrm{kin}} < 35$ K) relative to the observed line widths ($\sim$3 km s$^{-1}$) in \Gcloud\ and we ignore this correction.

\subsection{Column Density from MAMBO}

\begin{figure}
\includegraphics[width=0.48\textwidth]{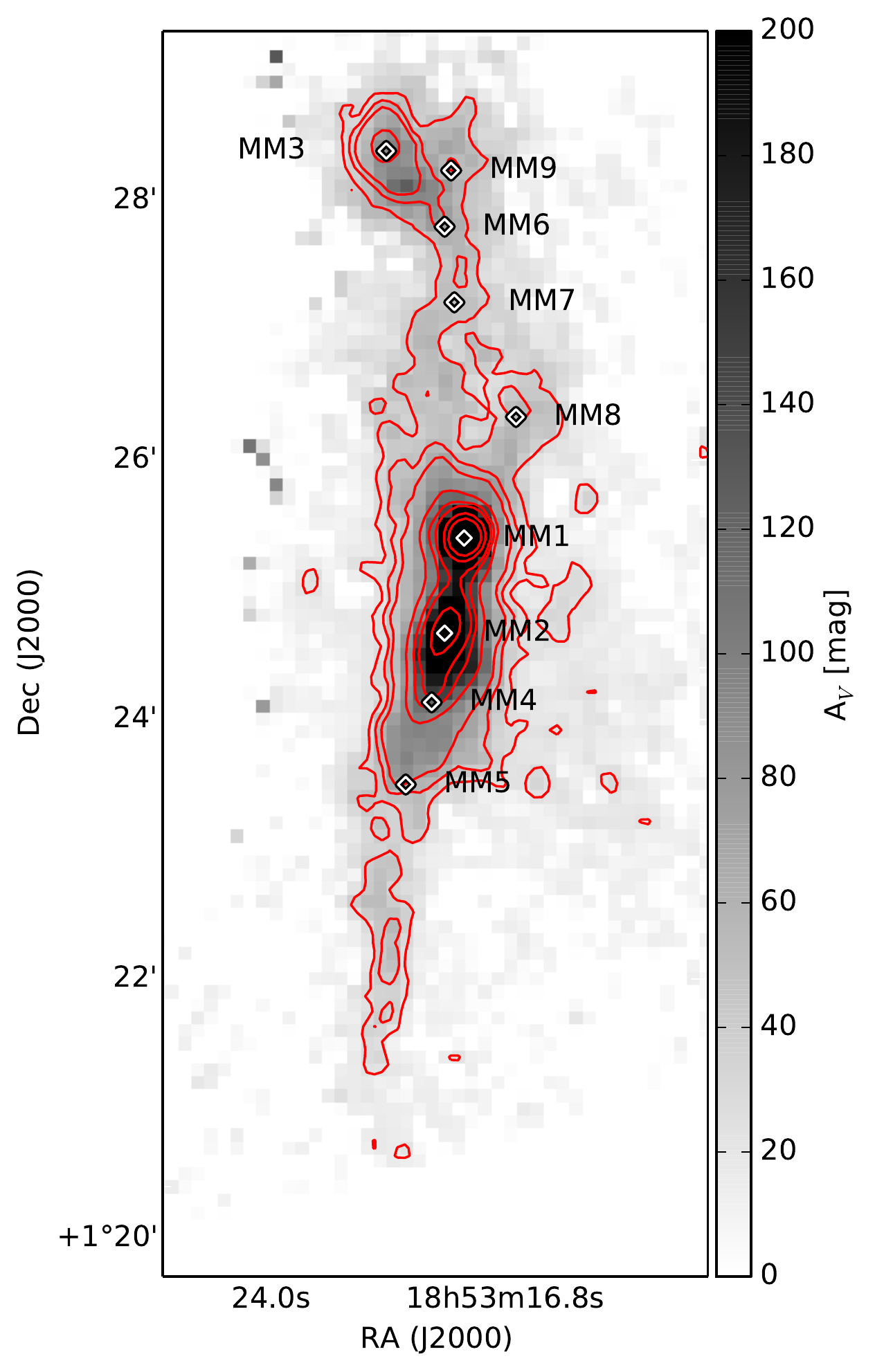}
\caption{Column density in magnitudes of visual extinction from combining the dust temperature map (\autoref{fig:dust_temp}) with the 1.2~mm fluxes from \autoref{fig:mambo_overview}. The same 1.2~mm contours and clumps as in \autoref{fig:mambo_overview} are shown for reference.}
\label{fig:column_density}
\end{figure}

Based on the preceding expectations about the biases in the two estimates of the dust temperature, we elect to use the $Herschel$ dust temperature for the remainder of the analysis. The column density from the 1.2 mm map is then calculated from the flux in each pixel from \autoref{eqn:bb}. For the MAMBO data $\lambda = 1200\, \mu$m, $\theta_{\mathrm{HPBW}} =$ 11\arcsec,  and $\kappa_{\nu} = 0.0102~$(cm$^{2}$ g$^{-1} $) \citep{Ossenkopf:1994}. This value of $\kappa_{\nu}$ corresponds to the wavelength-interpolated \citet{Ossenkopf:1994} value for dust grains with thin ice mantels in dense regions ($n$ = 10$^{6}$ cm$^{-3}$). For conversion to extinction we use
\begin{equation}
N_{\mathrm{H}_2} = 9.4\times 10^{20} \mathrm{cm}^{-2} (A_V/\mathrm{mag}),
\end{equation}
and
\begin{equation}
A_K = 0.118 A_V,
\end{equation}
from \citet{Bohlin:1978} and \citet{Rieke:1985}, respectively. This column density map is shown in \autoref{fig:column_density}; it preserves the high angular resolution of the MAMBO 1.2~mm data, subject to the twin assumptions that (1) the single-temperature dust-temperature fit to $Herschel$ and smoothed 1.2~mm data is a good first-order approximation of the average dust temperature along a line of sight and (2) the dust temperature of the bulk cloud is slowly varying (so that temperature variations occur on scales larger than our smoothed beam size). 

\subsection{Estimating the Dynamical State of the Filament}

\begin{figure}
\includegraphics[width=0.48\textwidth]{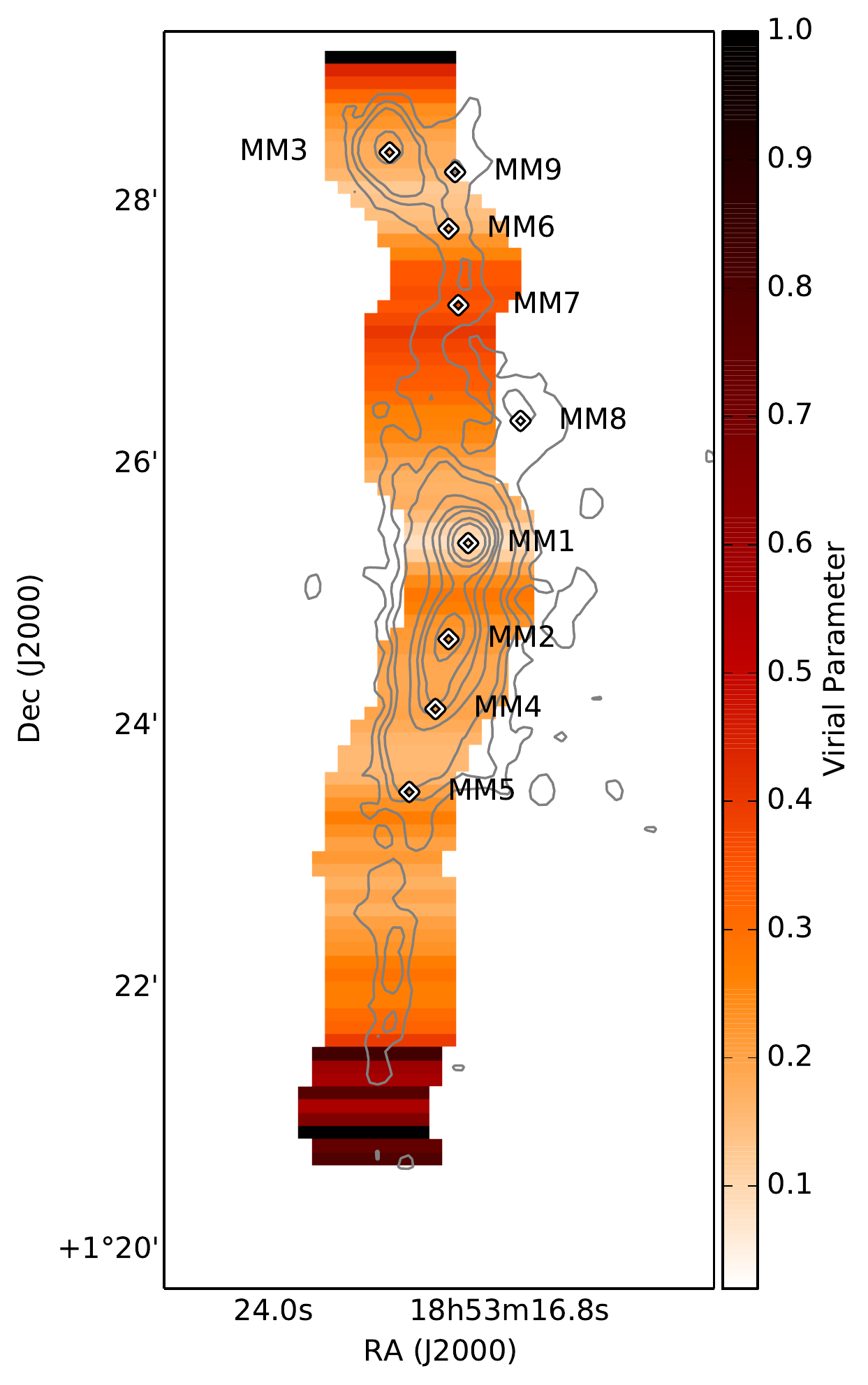}
\caption{Virial parameter calculated along the filament. For a cylinder this quantity is defined by integrating across the width of the cylinder, and so is calculated over a given length. This quantity is calculated per 0.11 pc (corresponding to one pixel in the NH$_3$ map). Regions with $\alpha$ $<$ 1 are nominally gravitationally bound, suggesting that this cloud is globally bound. The same 1.2~mm contours and clumps as in \autoref{fig:mambo_overview} are shown for reference.}
\label{fig:virial_per_area}
\end{figure}

The dynamical state of a self-gravitating filament can be estimated by comparing the linear mass density against the pressure support. This analysis ignores magnetic fields and the influence of external pressure, and assumes that the line width is due to turbulence rather than systematic motions like rotation or radial infall/expansion. An alternative approach is to assume that such filaments are in virial equilibrium and use this assumption to deduce the strength of magnetic support and external confinement \citep{Contreras:2013}. The pressure support is, in theory, a combination of thermal support and turbulent support; in the case of this cloud, however, thermal support is negligible compared to turbulent support and can be ignored. We characterize the dynamical state using the virial parameter $\alpha$ for an isothermal cylinder, which is, in useful units, 

\begin{equation}
\alpha = \frac{84 (\Delta V)^2}{(M/l)},
\end{equation}
where $M/l$ is the mass in \Msun\ per parsec and $\Delta V$ is the measured FWHM of a molecular line in units of km s$^{-1}$ \citep{Jackson:2010}.  For $\alpha <$ 1, the filament should collapse radially under its own mass (unless supported by, e.g., magnetic fields) and $\alpha >$ 1 implies that the filament is confined (perhaps by external pressure) or is destined to disperse. For the regions that have already formed protostars, the observed line width may be due to protostellar outflows keeping the gas close to virial equilibrium  \citep{Li:2006}.

The mass per unit length of the filament comes from summing the column density map shown in \autoref{fig:column_density}. The line width map (\autoref{fig:linewidth}) comes from fitting the NH$_3$ data. The column density map traces the main filament (note that this excludes MM8, a more isolated clump composed of two distinct velocity components). We calculate the virial parameter along the length of the filament by adopting a fixed length step of 11\arcsec, the resolution of our column density map, and using the median $\Delta V$ in this portion of the filament from the NH$_3$ map. The NH$_3$ map from which we measure $\Delta V$ is at lower resolution (31\arcsec), so adjacent slices are not independent. This map is shown in \autoref{fig:virial_per_area}. The majority of the cloud has $\alpha < 1$, except at the ends of the filament, suggesting that the cloud is globally bound. 

\section{Identification of Protostars}
\label{sec:identification}

\subsection{Extended $K$-band Sources}
\capstartfalse
\begin{deluxetable}{lccccc}
\tablewidth{0pc}
\tablecaption{Extended $K$ band Sources}
\tablehead{
\colhead{Obj.} & \colhead{R.A.} & \colhead{Decl.} & \colhead{Field} & \colhead{$K$} & \colhead{$H$-$K$}\\
	& hh:mm:ss & dd:mm:ss & & &
}
\startdata
K1 & 18:53:16.99 & 1:26:41.28 & 1 & 17.646$\pm$0.014 & 3.26$\pm$0.06\\
K2 & 18:53:17.35 & 1:26:33.00 & 1 & 15.666$\pm$0.002 & 2.43$\pm$0.01\\
K3 & 18:53:19.66 & 1:28:16.68 & 6 & $>$21 & N/A\\
K4 & 18:53:19.92 & 1:28:08.04 & 6 & 19.075$\pm$0.022 & $>$6\\
K5 & 18:53:20.02 & 1:28:19.56 & 6 & 17.123$\pm$0.004 & 1.14$\pm$0.01\\
K6 & 18:53:20.02 & 1:28:27.84 & 6 & 18.722$\pm$0.016 & 3.70$\pm$0.20\\
K7 & 18:53:20.04 & 1:28:14.16 & 6 & 18.660$\pm$0.014 & 1.52$\pm$0.03\\
K8 & 18:53:20.06 & 1:28:23.88 & 6 & $>$21 & N/A\\
K9 & 18:53:20.50 & 1:28:36.84 & 6 & 18.369$\pm$0.010 & 1.24$\pm$0.01\\

\enddata
\label{table:extended_sources}
\tablecomments{$K$-band magnitudes and $H-K$ colors are listed for the point source (1\arcsec\ FHWM) photometry performed on the central position of the source. A limiting magnitude for the field is given where the point source is not detected.}
\end{deluxetable}
\capstarttrue

Objects with $K$-band nebulosity are highly likely to be embedded sources and therefore protostars. Although not all young stars will be identified in this manner, the presence of these sources definitively identifies young stars and the correspondence of these sources with the regions identified statistically (see \autoref{sec:nir}) provides an important independent validation of this statistical method.

\begin{figure}
\includegraphics[width=0.48\textwidth]{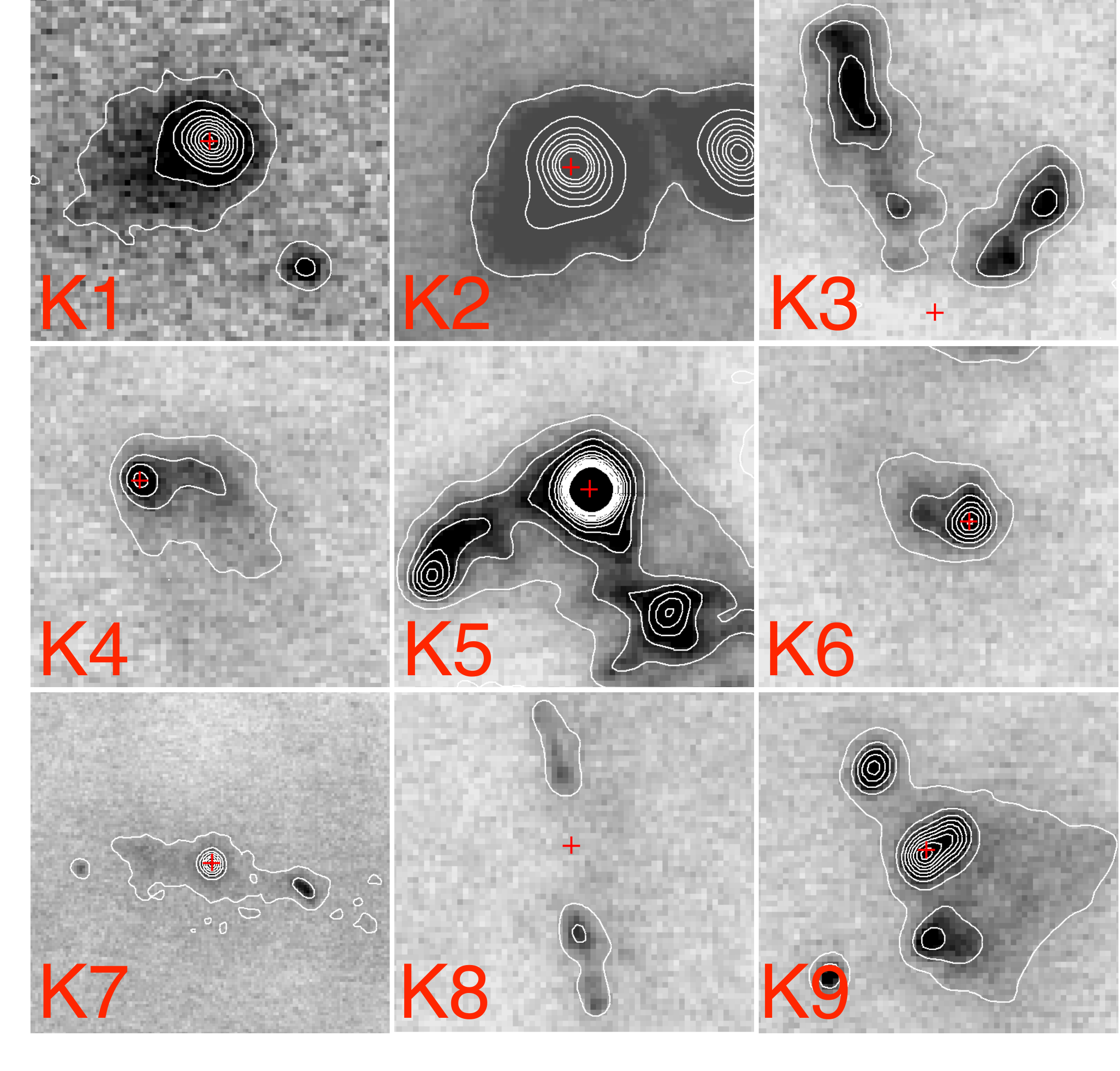}
\caption{The nine most significant extended $K$-band sources, as given in \autoref{table:extended_sources}. Of these sources, K1 and K2 lie in Field 1 while the remainder lie in Field 6. The red cross shows the location of the source listed in \autoref{table:extended_sources}. For sources K3 and K8, which are plausibly outflow lobes from an unseen central source, we have assigned a possible location for the driving source. All fields are 2\arcsec.5 on a side, except for K7, which is 5\arcsec on a side to accommodate its extent.}
\label{fig:extended_sources}
\end{figure}

\begin{figure}
\includegraphics[width=0.48\textwidth]{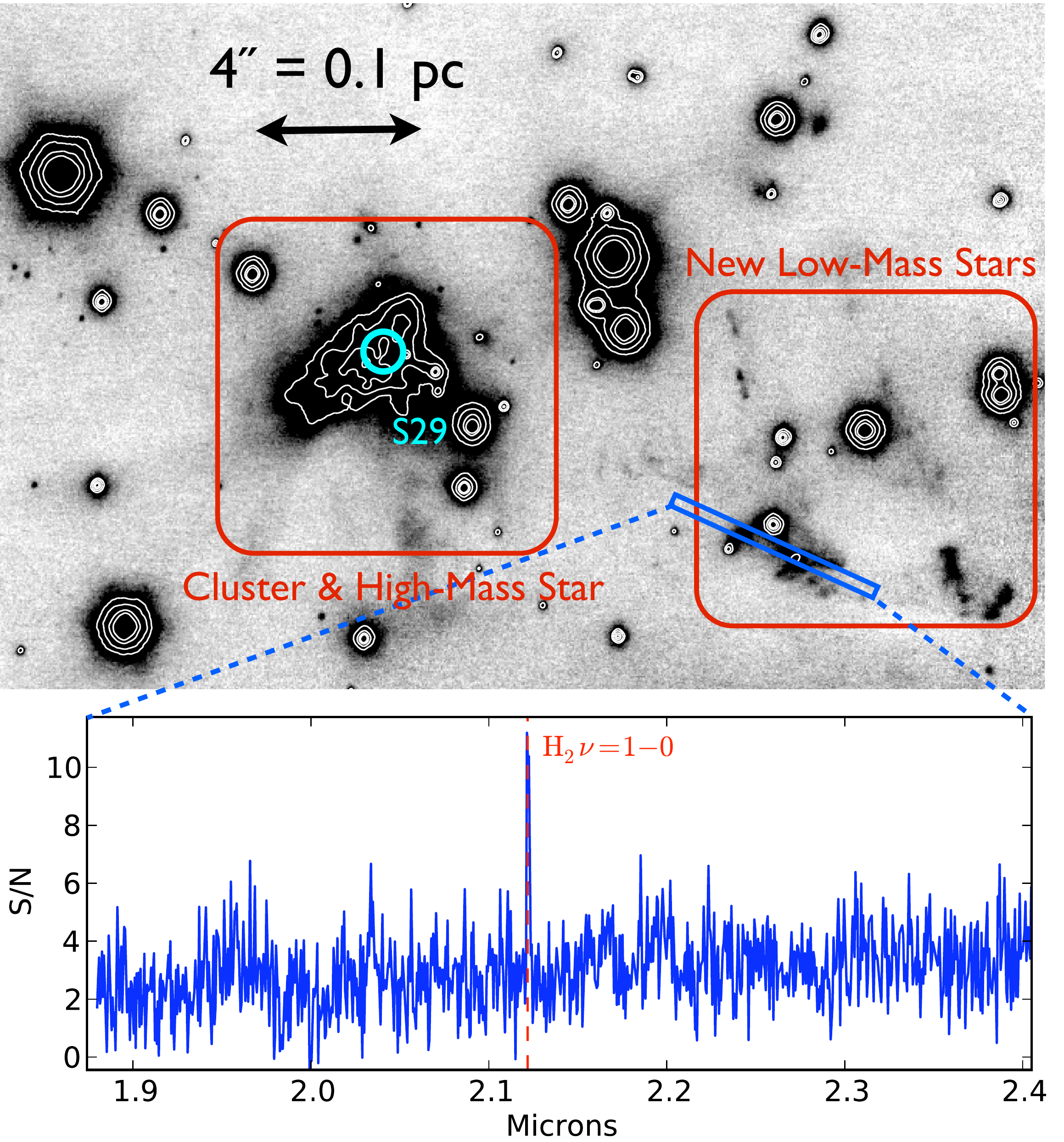}
\caption{Portion of \mmthree\ with the $K$-band image in grayscale. The combination of outflows and point sources around \citet{Shepherd:2007} source no. 29 (S29) were unresolved in \citet{Shepherd:2007} and compose a small cluster. A $K$-band spectrum of the brightest of the new extended $K$-band sources shown in \autoref{fig:extended_sources} (K5) shows that the emission is dominated by H$_2$ emission, most likely arising from shocked or UV excited gas in a protostellar outflow. }
\label{fig:irtf_spectra}
\end{figure}

$K$-band nebulosity can arise from protostellar outflows or reflection nebulosity \citep{Connelley:2007}. In \autoref{fig:extended_sources} we show the nine most clearly extended new sources in our Keck images. Seven of these appear to be a point source with some extended nebulosity, while the other two features appear morphologically similar to protostellar outflows around deeply embedded and unseen powering sources. In addition to these nine new sources, \citet{Shepherd:2007} source no. 29 (henceforth S29, see \autoref{fig:irtf_spectra}) appears as an extended source in the $K$ band in UKIDSS.

We obtained SPeX spectra of the two brightest extended $K$-band sources in Field 6 (the only field imaged before the spectroscopic observations), K5 and S29. The spectra from these sources (\autoref{fig:irtf_spectra}) shows that the $K$-band emission is dominated by the H$_2$ $\nu$ = 1-0 emission line, which is indicative of excited H$_2$, most likely from a protostellar outflow or UV excitation within an outflow cavity. The large extended $K$-band nebulosity around S29 \citep[Peak A from][]{Sakai:2013} is also dominated by H$_2$ emission, with some continuum point sources embedded. The complicated morphology of this nebulosity suggests that this is a cluster of many protostars with the most massive member, presumably, being Peak A from \citet{Sakai:2013}. The remaining extended $K$-band sources in Field 6 were too faint for useful spectroscopic observations. 

\subsection{Overdensities of Red Sources}
\label{sec:nir}
We identify low-mass protostars by looking for a statistical excess of red sources. \Gcloud\ has a high column density, so over most of the cloud, background stars can be separated from foreground stars with a color cut in $H-K$. In the case of a cluster of protostars, there will be an excess of red sources over what is expected. The expected number of background stars at a given observed magnitude depends on the column density of the cloud at that point.

We use the Besan\c{c}on stellar model \citep{Robin:2003} to estimate the spatial density of background sources as a function of $K$-band magnitude. We include 1.5 mag of visual extinction per kpc as the diffuse extinction component in this model. This model is useful because the UKIDSS data is not deep enough (the catalog is complete to $K$ = 18 mag) and the Keck images are positioned on the cloud. The UKIDSS and Keck data can be used to verify the reliability of the model source counts at the bright ($K < 17$ mag) and faint ($17 < K < 20$ mag) ends respectively. Due to the narrow range of intrinsic stellar $H-K$ colors, in sufficiently dense portions of the cloud ($A_V > 10$ mag), the color distribution is bimodal, and one can estimate the number density of foreground (i.e., blue) stars as a function of $K$-band magnitude and subtract this density from the $K$-band magnitude distribution observed off the cloud. In this case, off the cloud means within one square degree of the cloud and more than 2\arcmin\ from the regions of significant column density as shown in \autoref{fig:column_density}. This provides an estimate of the spatial number density of background sources as a function of their observed $K$-band magnitudes.

\begin{figure}
\includegraphics[width=0.48\textwidth]{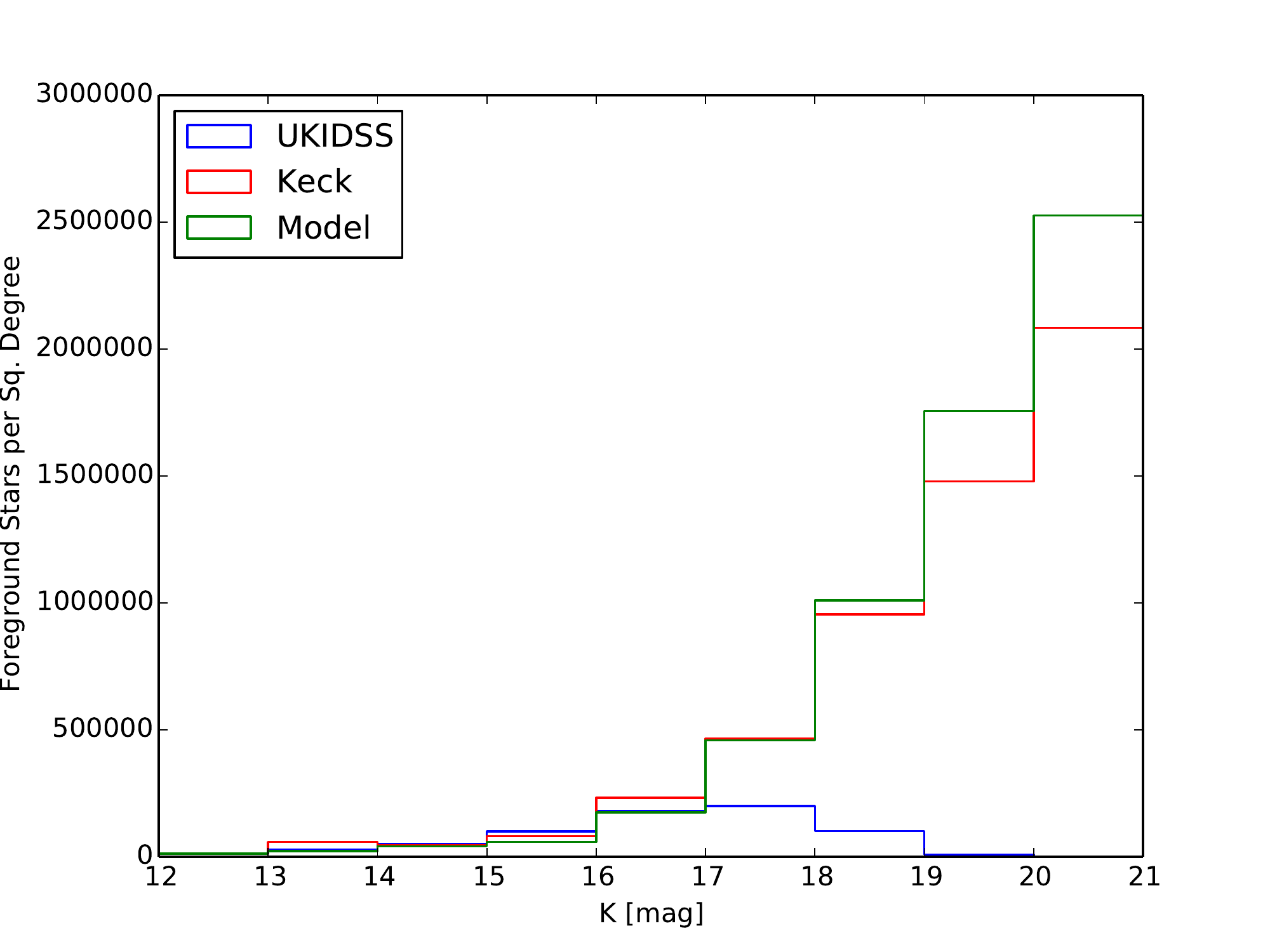}
\caption{Estimated spatial number density of background sources as a function of observed $K$-band luminosity for the Besan\c{c}on model (green) and UKIDSS (blue) and Keck (red) data sets. }
\label{fig:Khist}
\end{figure}

Comparison to the Besan\c{c}on model shows good agreement between the number of stars expected and observed. As shown in \autoref{fig:Khist}, there is some divergence at the faintest end (in the sense that the model predicts more faint stars than we observe). This is likely due to the simple treatment of extinction in the model. However, this deficiency is relatively unimportant. For the dense portions of the cloud, only background stars that are intrinsically very bright will be observed. Getting the faint end of this distribution correct is really only important for preventing spurious signals at the low-density edge of the cloud. Examination of the model colors as well as observations of the bimodal distribution of $H-K$ colors on dense portions of the filament show that adopting a limit of $H-K > 2$ mag is a good conservative estimate for distinguishing foreground from background sources, as $>99\%$ of all un-reddened stars have $H-K < 2$ mag \citep{Lucas:2008}. Note that there are some sources with $H-K > 2$ mag far from the filament in the UKIDSS data (see \autoref{fig:cmd}), but these sources are reasonably explained as stars behind the other less dense dust features present in this part of the Galactic plane and the rare asymptotic giant branch star with intrinsically red colors. 

\begin{figure*}
\centering
\includegraphics[height=0.86\textheight]{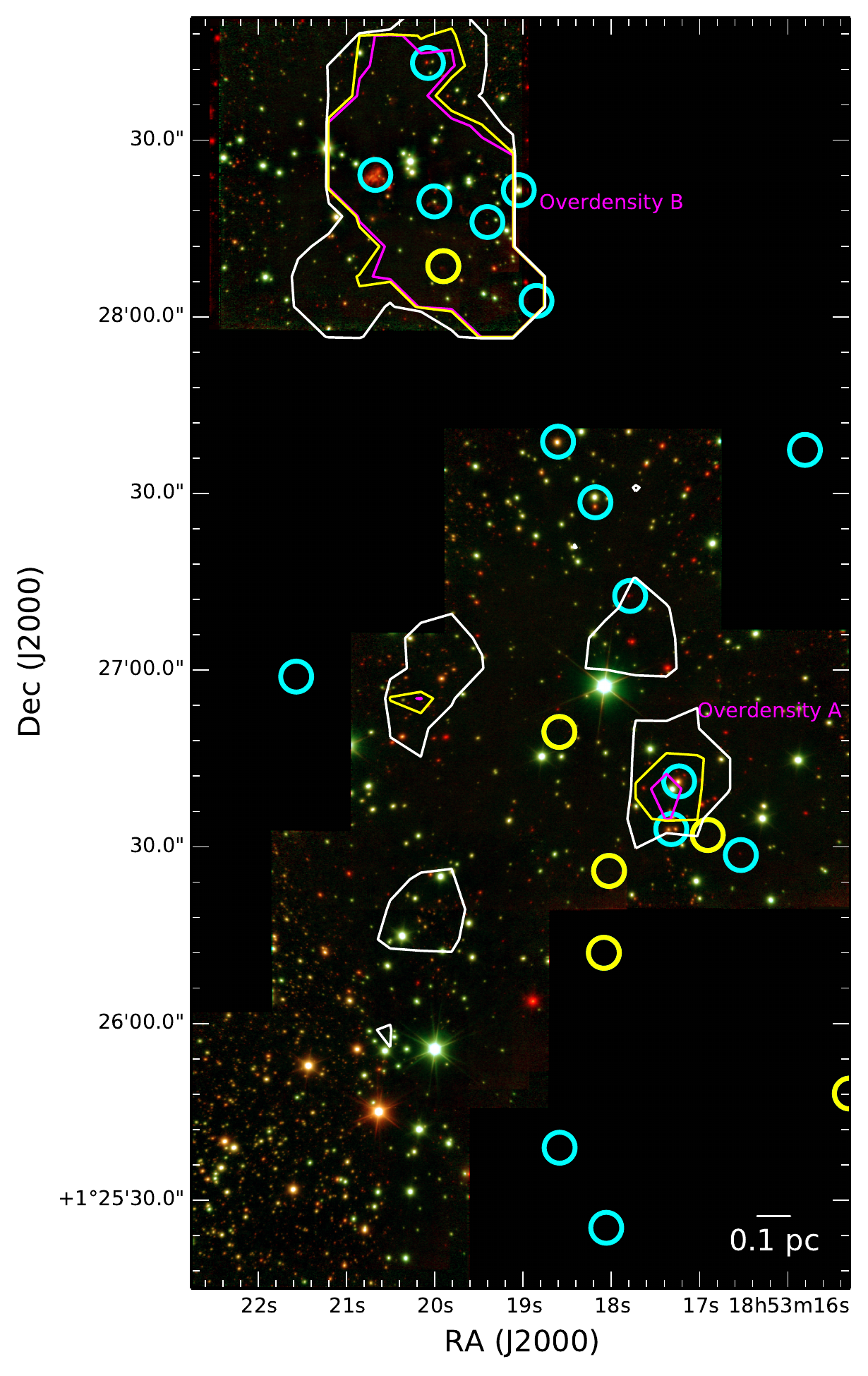}
\caption{Two-color image of Fields 1 to 6 ($K$:red, $H$:green) with contours showing how likely it is that the number of red sources seen at a given location is due to background stars as derived from Keck data. Contours are at probabilities of 0.005, 0.01, and 0.05 with magenta, yellow, and white contours, respectively. Thus, the red source density of the region within the magenta contour is less than 0.5\% likely to be due to background stars shining through the column density of the cloud at this location. Candidate YSOs from \citet{Shepherd:2007} are shown in cyan (sources with a spectral energy distribution fit) and yellow (24 \micron\ point sources without a full fit) circles.}
\label{fig:1to5}
\end{figure*}

\begin{figure}
\centering
\includegraphics[width=0.48\textwidth]{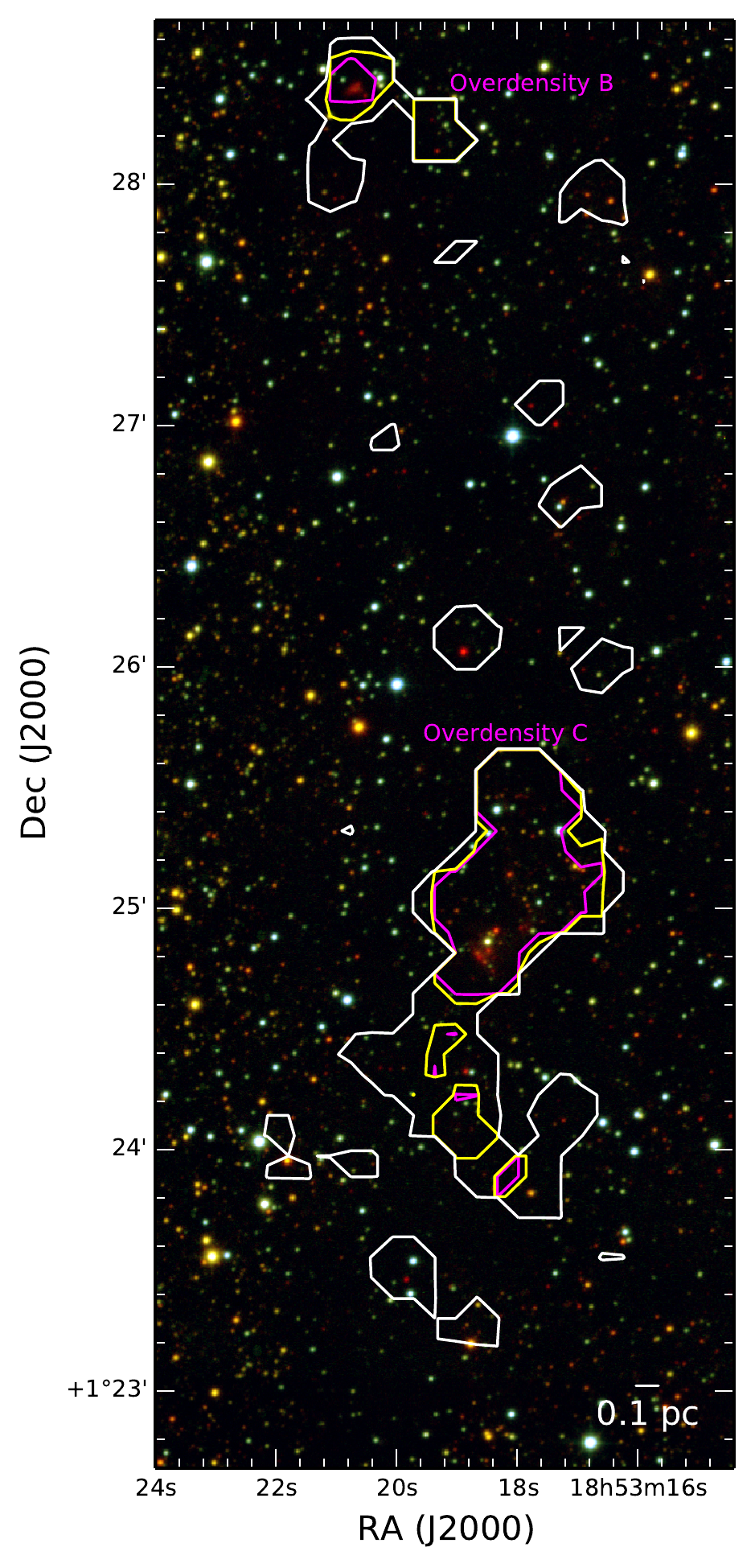}
\caption{Three-color image of the full filament ($K$:red, $H$:green, and $J$: blue) with contours showing how likely it is that the number of red sources seen at a given location is due to background stars as derived from UKIDSS. Contours are at probabilities of 0.005, 0.01, and 0.05 with magenta, yellow, and white contours, respectively. The fainter limiting magnitude of UKIDSS produces a smaller number of red background and/or embedded sources than is possible with Keck.}
\label{fig:whole_cloud}
\end{figure}

With the extinction map shown in \autoref{fig:column_density}, we can correct the observed $K$ magnitudes for all red stars to intrinsic magnitudes, which we denote as $K'$. Given a sample of $N$ red stars in an area with intrinsic (un-reddened) $K$-band magnitudes, \{$K'_1$, $K'_2$, \dots, $K'_N$\}, we can ask, what is the probability, upon drawing a random sample of the appropriate size from the Besan\c{c}on model (a model draw), that the cumulative magnitude distribution of the model draw is strictly greater than or equal to the cumulative magnitude distribution of the observed stars? That is, what is the chance that the brightest star from the model draw is brighter than the brightest observed stars, and the $n$th brightest star from the model draw is brighter than the $n$th brightest observed star for all $N$ stars? This random drawing process is computationally rather slow, and gives no information about the relative likelihood that a given star is background versus non-background.

Instead, one can model this process as follows. The Besan\c{c}on model gives, for a given $K'$ magnitude, the expected surface density of background stars at least as bright as $K'$, $\rho(K \le K')$. Now if we consider the observation of one red star at a given $K'_i$, the expected number of stars at least this bright in an area $\Omega$ is $\lambda = \rho(K \le K') \Omega$, and the number of stars $k$ within an area of $\Omega$ is Poisson distributed with expected value $\lambda$ such that
\begin{equation}
P(n=k) = \frac{\lambda^k e^{-\lambda}}{k!},
\end{equation}
and the probability of seeing at least one star in $\Omega$ is 
\begin{equation}
P(n \ge 1) = 1-P(n=0). 
\label{eqn:p}
\end{equation}

For a given area, $\Omega$, wherein we observe $N$ stars with different intrinsic (unreddened) magnitudes $K'_i$, the probability of seeing at least this many stars is less than 
\begin{equation}
P(N) = \prod_{i} (1-P(n=0)).
\label{eqn:fullprob}
\end{equation}
Note that this calculation uses $P$($n \geq 1$) since the color cut is conservative and may not include all background stars. Where this quantity is very small, the distribution of sources cannot be explained as being due to background stars and so we assume it arises from a cluster of protostars. We define significance as less than a 0.5\% chance that the red sources in this region could be reproduced by background stars. 

\autoref{eqn:fullprob} is only an upper bound on the probability. Consider the case of a region with just two red stars; a bright and a faint star with magnitudes $K'_1$ and $K'_2$, respectively. The probability in \autoref{eqn:p} is appropriate for the brighter star, but for the fainter star, there are two stars at least as bright as $K'_2$, so \autoref{eqn:p} should be replaced by $P(n \ge 2) = 1-P(n=0)-P(n=1)$. Ignoring this complication allows one to calculate a single probability for each red star from \autoref{eqn:p}, without considering the influence that other stars will have in a particular resolution element. In addition, \autoref{eqn:fullprob} ends up down-weighting the statistical influence of the faint end of the magnitude distribution. This is actually desirable, as the density of sources from the Besan\c{c}on model is less certain at the faint end (\autoref{fig:Khist}). Thus, by adopting this approximation, the process is less sensitive to real overdensities, but less likely to generate regions of false significance due to incorrectly modeling the faint end of the magnitude distribution. 

We make a map of the significance of the excess red sources over each field by stepping through each field and assessing the significance within a radius of 11\arcsec. This radius was chosen to match the resolution of the column density/extinction map.

The other critical consideration is completeness. We consider sources that are above the completeness limit in $K$ for each field and at least 2 mag brighter than the $H$-band completeness limit (so that a non-detection in the $H$ band is still a genuinely red source with $H-K >$ 2). We apply the same analysis to the UKIDSS data. The fainter limiting magnitude in the UKIDSS data means that it is less sensitive in identifying protostars. The limiting magnitudes used are given in \autoref{table:fields}.

The results of this analysis are shown in \autoref{fig:1to5} and \autoref{fig:whole_cloud}. Field 7 is not shown, but shows no areas of significant probability. Out of the seven Keck fields, two have regions where there is a significant increase in sources over the expected background density. Note that these regions do not necessarily correspond to regions with the greatest density of red sources, as the column density of the cloud (and hence the diminution of background stars) changes. The two fields with overdensities are Fields 1 and 6, near MM8 and MM3, respectively. The overdensities identified in UKIDSS are near MM3 (i.e., a smaller region than already identified in Keck Field 6) and covering MM1 and MM2. These overdensities are listed in \autoref{table:clusters}. Because this is a statistical determination, it is not possible to say with great confidence that any given source is a protostar. \autoref{fig:cmd} highlights individual sources with $P$($X_i=k$) $<$ 0.25, but it is only the spatial coincidence of many such sources that allows us to identify an overdensity.

\capstartfalse
\begin{deluxetable}{lccccc}
\tablewidth{0pc}
\tablecaption{Near-infrared Source Overdensities}
\tablehead{
\colhead{Name} & \colhead{R.A.} & \colhead{Decl.} & \colhead{$\Delta$R.A.} &  \colhead{$\Delta$Decl.} & \colhead{Field} \\
			 &  \colhead{hh:mm:ss}  & \colhead{dd:mm:ss} & & &}
\startdata

 A & 18:53:20.50 & 1:26:37.32 & 22\arcsec & 30\arcsec & Keck Field 1 \\
 B & 18:53:20.04 & 1:28:19.20 & 48\arcsec & 68\arcsec & Keck Field 6 \\
 C & 18:53:18.19 & 1:25:06.96 & 36\arcsec & 76\arcsec & UKIDSS \\

\enddata
\label{table:clusters}
\tablecomments{ $\Delta$R.A. and $\Delta$Decl. list the approximate size of the region over which there is a significant overdensity of red sources. }
\end{deluxetable}
\capstarttrue

\section{Discussion}

\subsection{Identification of Low-mass Protostars}
\label{subsec:identification}
Both the regions with overdensities of red sources identified in the Keck images are independently confirmed as containing protostars by the presence of extended $K$-band sources since it is precisely these fields that contain sources K1 - K9 (see \autoref{fig:extended_sources} and \autoref{table:extended_sources}).

The near-infrared identification of low-mass protostars is clearly highly selective. Large extinction can render a protostar invisible in the $K$ band to arbitrary depths. Near-infrared identification is complementary to other approaches. The main downside to space-based mid-infrared searches (i.e., $Spitzer$ and $Herschel$) is that the low-resolution and large contrast ratios between low- and high-mass protostars in the mid-infrared mean that it is very difficult to find low-mass protostars in the vicinity of high-mass protostars. Conversely, in the near-infrared, the contrast ratio in stars of different mass is lower because the observed flux is a strong function of the extinction rather than the intrinsic luminosity (unfortunately this makes it basically impossible to estimate a mass for a protostar simply from the near-infrared magnitudes). 

\begin{figure}
\includegraphics[width=0.48\textwidth]{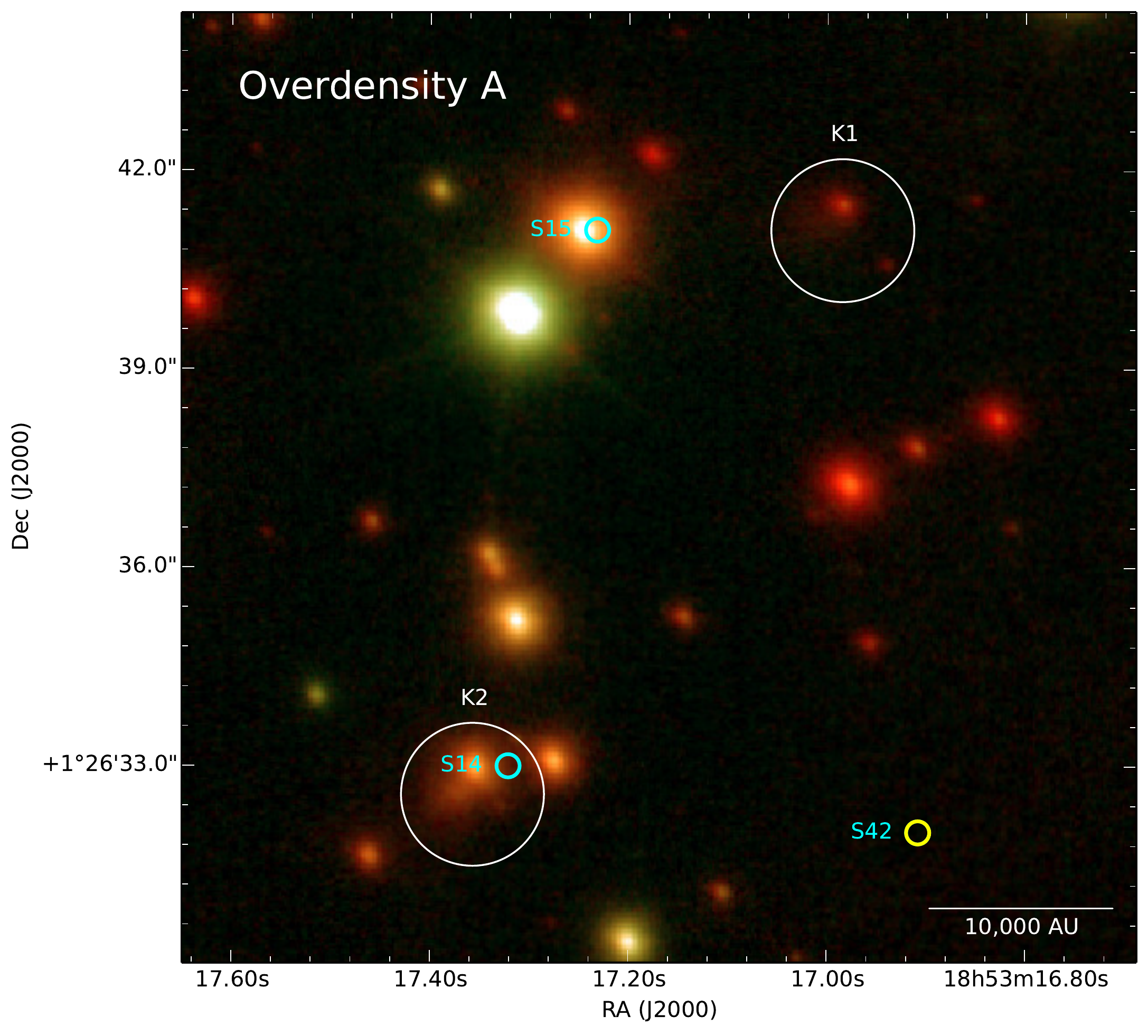}
\caption{Portion of \fonethreeone\ with an excess of red sources. Two-color image with $K$ in red and $H$ in green. At least two of these sources are clearly extended. These could be either wide-angle outflows or near-infrared reflection nebulosity. In either case, this suggests that they are young sources associated with the cloud. Sources from \citet{Shepherd:2007} are labeled as S14, S15, and S42.}
\label{fig:f13_1_zoom}
\end{figure}

The advantage of deep near-infrared images is shown in \autoref{fig:f13_1_zoom}, which shows Overdensity A in the near-infrared along with the three candidate protostars (identified with $Spitzer$) from \citet{Shepherd:2007}. The higher resolution near-infrared data shows that S14 is composed of two sources. One of these is an extended source with $K$-band nebulosity, suggesting that this one (at least) is a genuine protostar. The other extended $K$-band source in this region is not identified as a candidate protostar in \citet{Shepherd:2007}; its mid-infrared flux is hard to estimate due to the presence of bright emission from the nearby source S15. Finally, S42 from \citet{Shepherd:2007} is only detected at 24~\micron. Even in these deeper images we do not see the source in the near-infrared; this does not significantly constrain whether this source is a deeply embedded protostar or a background 24~\micron\ source.

\begin{figure}
\includegraphics[width=0.480\textwidth]{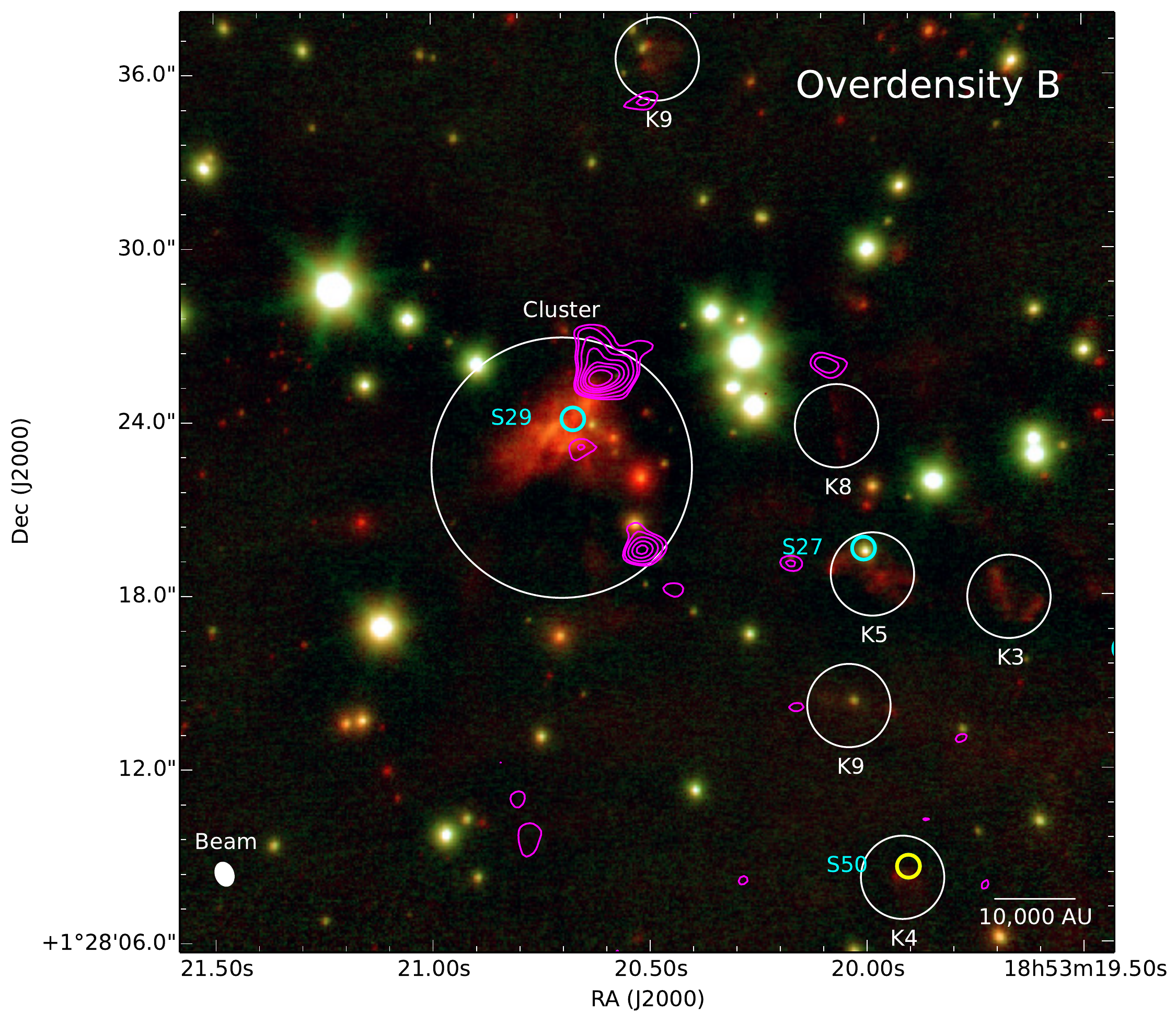}
\caption{Portion of \mmthree\ showing low-mass protostars as seen in the near-infrared and millimeter. Two color image with $K$ in red and $H$ in green. Magenta contours show ALMA continuum at 1.3 mm with contours at 1.5, 2.0, 3.0, 4.5, 6.0, 7.5, and 10 mJy beam$^{-1}$ and the beam shown in the lower-left. The extended $K$-band sources with outflows or reflection nebula (white circles) reveal a less embedded (presumably older) population of low-mass protostars. Many ALMA continuum sources remain unseen in the $K$ band. Sources from \citet{Shepherd:2007} are labeled as S27, S29, and S50. }
\label{fig:mm3_zoom}
\end{figure}

Both mid- and near-infrared observations may miss the earliest, most deeply embedded phase of protostars. In \autoref{fig:mm3_zoom} we show ALMA continuum at 1.3 mm observations around Overdensity B/MM3. \citet{Sakai:2013} identified a hot core (or hot corino, depending on the distance assumed for \Gcloud) associated with the strongest millimeter source seen in the ALMA observations. This corresponds to an unseen source at the northern edge of the complicated cluster of protostars and outflows seen in the $K$ band. Also present within the image are a number of other millimeter continuum detections of varying significance. Several of the strong sources are near extended $K$-band sources, particularly K8 and K9. In particular in K8, this could point to the location of the Class 0 source powering the near-infrared outflow. The primary beam of the ALMA observations is larger than the region shown in \autoref{fig:mm3_zoom}, but there are no continuum sources at this flux level east of the main cluster; all the continuum sources are to the west, spatially coincident with the overdensity of red near-infrared sources.

\subsection{Low-mass Stars Forming First?}

There is some debate about how star formation proceeds within a filamentary molecular cloud. \citet{Jackson:2010} has proposed that the ``sausage'' instability produces relatively evenly spaced dense molecular clumps, and that star formation, particularly high-mass star formation takes place within those. \Gcloud\ seems to fit within this paradigm, as it exhibits a series of dense molecular clumps \citep[as identified by][]{Rathborne:2006} with relatively even spacing (0.74$\pm$0.08 pc). Some of these clumps are forming stars, with some of the most massive, MM1, MM2,  MM3, and (probably) MM4 currently forming high-mass stars \citep[][and references therein]{Rathborne:2005, Chambers:2009, Sanhueza:2010}. 

The remaining millimeter clumps show less star formation activity. One explanation is that they are generally less massive (90 - 160 \Msun). A simple estimate suggests that a clump mass of roughly 200 \Msun\ is necessary to form a high-mass star and associated cluster \citep{Jackson:2013}. It is therefore possible that the remaining millimeter clumps will not form high-mass stars. Indeed, for MM8 the NH$_3$ observations show the presence of two velocity components toward this core separated by 2 km s$^{-1}$; MM8 is likely the superposition of two less massive clumps. Alternatively, given at least a factor of two uncertainty in the determinations of these clump masses and the possibility that they will continue to accrete mass from the filament, these currently starless clumps could still be the site of future high-mass star formation. 

The new low-mass protostars in this study are both situated between existing clumps (Overdensity A lies between MM7 and MM8, and Overdensity B lies between MM3 and MM9). Our ability to detect low-mass protostars in the near-infrared is diminished within the high-extinction millimeter clumps. However, we have shown that within this young cloud there are already low-mass protostars forming in the inter-clump medium of a filamentary molecular cloud. 

\Gcloud\ is a single coherent molecular cloud in the sense that the range of NH$_3$ centroid velocities seen in the filament is small and varies smoothly across the cloud. Therefore, it makes sense to talk about the relative formation times for low-mass stars and high-mass stars within the whole cloud. On a smaller scale, we might want to know if a given population of low-mass stars is close enough to a site of potential high-mass star formation to end up influencing future star formation. In the vicinity of Overdensity B and MM3, the typical dust temperature is 20 K and the typical volume density (assuming that the projected diameter of the filament is equal to its width) is 3.2$\times10^4$ cm$^{-3}$. This leads to a Jeans length \citep{Stahler:2004} of $\lambda_J$ = 0.15 pc or 3.1$\times10^4$ AU. From \autoref{fig:mm3_zoom} it is therefore clear that the newly identified low-mass stars are a few Jeans lengths away from the central cluster, and therefore are probably not immediately relevant to the formation of a high-mass star in the central cluster.

\citet{Krumholz:2008} have proposed a critical column density threshold of 0.7-1.5 g cm$^{-2}$ in order to form massive stars. In that work, this is the critical column density at which the number of low-mass protostars that form within a region is large enough that their radiation suppresses fragmentation enough to form stars with masses $>$ 10 \Msun. Our estimated column densities in MM3 and Overdensity B are roughly 0.1 g cm$^{-2}$, and so by this calculation the region is not dense enough for the low-mass stars to be able to have a significant ability to raise the ambient temperature of the cloud. 

Both these lines of evidence suggest that, although on the cloud scale, low-mass stars may form at the same time or coevally with high-mass stars, these stars we are detecting are not likely to have a substantial impact on the high-mass stars that may form in the nearby dense clumps.

\subsection{Origin of Diffuse SiO Emission}
Diffuse, narrow line width SiO emission has been seen in several IRDCs \citep{Sanhueza:2013, Jimenez-Serra:2010}, including in regions with no known protostars. Since SiO emission in star-forming regions is normally associated with shocks from protostellar outflows, this emission is rather surprising. One possibility is that this SiO emission does arise from protostellar outflows, but that these outflows come from low-mass protostars that have hitherto escaped detection in these IRDCs. This narrow line width SiO component is seen in some regions with low-mass protostars and outflows, such as NGC 1333 \citep{Lefloch:1998}. The presence of a distributed population of low-mass protostars with signs of outflow activity in \Gcloud\ lends credence to this possibility, although it remains to be seen whether there are low-mass protostars with outflows actually present in the clouds with diffuse, narrow line width SiO emission.

\subsection{Distributed versus Clustered Star Formation}

\begin{figure}
\includegraphics[width=0.48\textwidth]{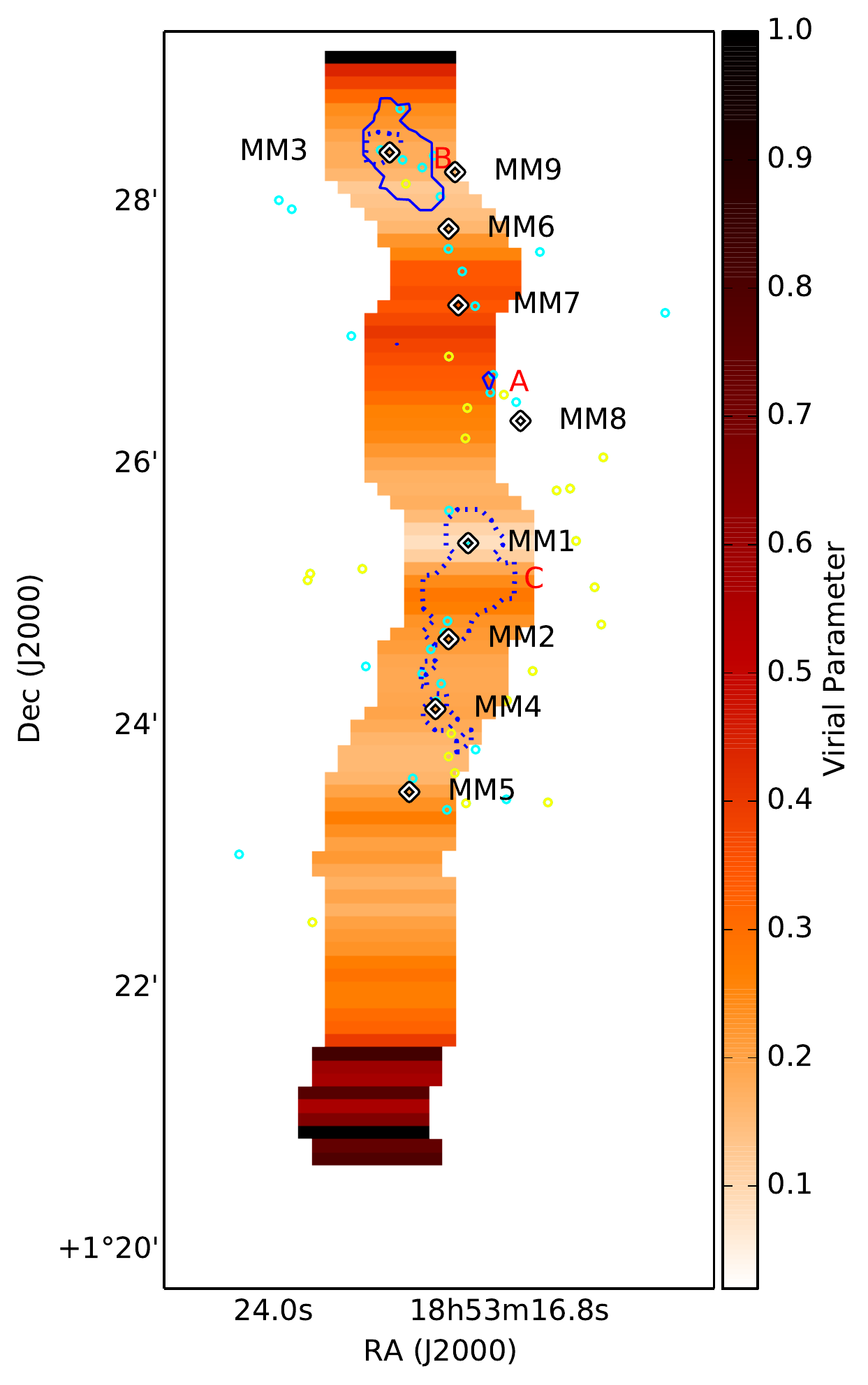}
\caption{Overdensities of protostars identified in \autoref{sec:identification} (blue contours) and candidate protostars from \citet{Shepherd:2007} (cyan and yellow points) shown on the virial-parameter map of \Gcloud. Overdensities are labled as in \autoref{table:clusters} and comprise those identified from Keck observations (solid) and UKIDSS (dashed). The same 1.2 mm clumps as in \autoref{fig:mambo_overview} are shown for reference.}
\label{fig:virial_per_area_stars}
\end{figure}

\citet{Bonnell:2011} model a filamentary cloud wherein some portions are bound and others are unbound. The bound regions have a higher star formation efficiency (up to 40\%) than the unbound regions (1\%). Furthermore, the IMF only matches the \citet{Chabrier:2003} IMF in the bound portions of the filament; in the unbound regions the IMF is strongly peaked at the local Jeans mass. This simulation did not include stellar feedback (either radiative feedback or outflows), which has been shown in other simulations to have a significant impact on the IMF \citep[i.e.][]{Krumholz:2011, Offner:2009}. 

The analysis presented here suggests that the entire \Gcloud\ cloud is gravitationally bound. However, there are significant assumptions underpinning the virial estimate. The emissivity, $\kappa$, and hence the mass, is uncertain by at least a factor of two. Our analysis has neglected significant possible extra energy sources in the cloud, including external (pressure) confinement and magnetic support. Our calculation of the mass depends on the distance to \Gcloud, which, as discussed in \autoref{sec:intro}, has conflicting estimates. Many of these assumptions introduce the possibility of systematic bias; for instance, changing the distance to \Gcloud\ would scale all masses. For this reason, the $relative$ boundedness of different portions of the filament is more reliably determined than the absolute value. \autoref{fig:virial_per_area_stars} shows the virial parameter along the filament with candidate protostars and identified overdensities.

\begin{figure}
\includegraphics[width=0.53\textwidth]{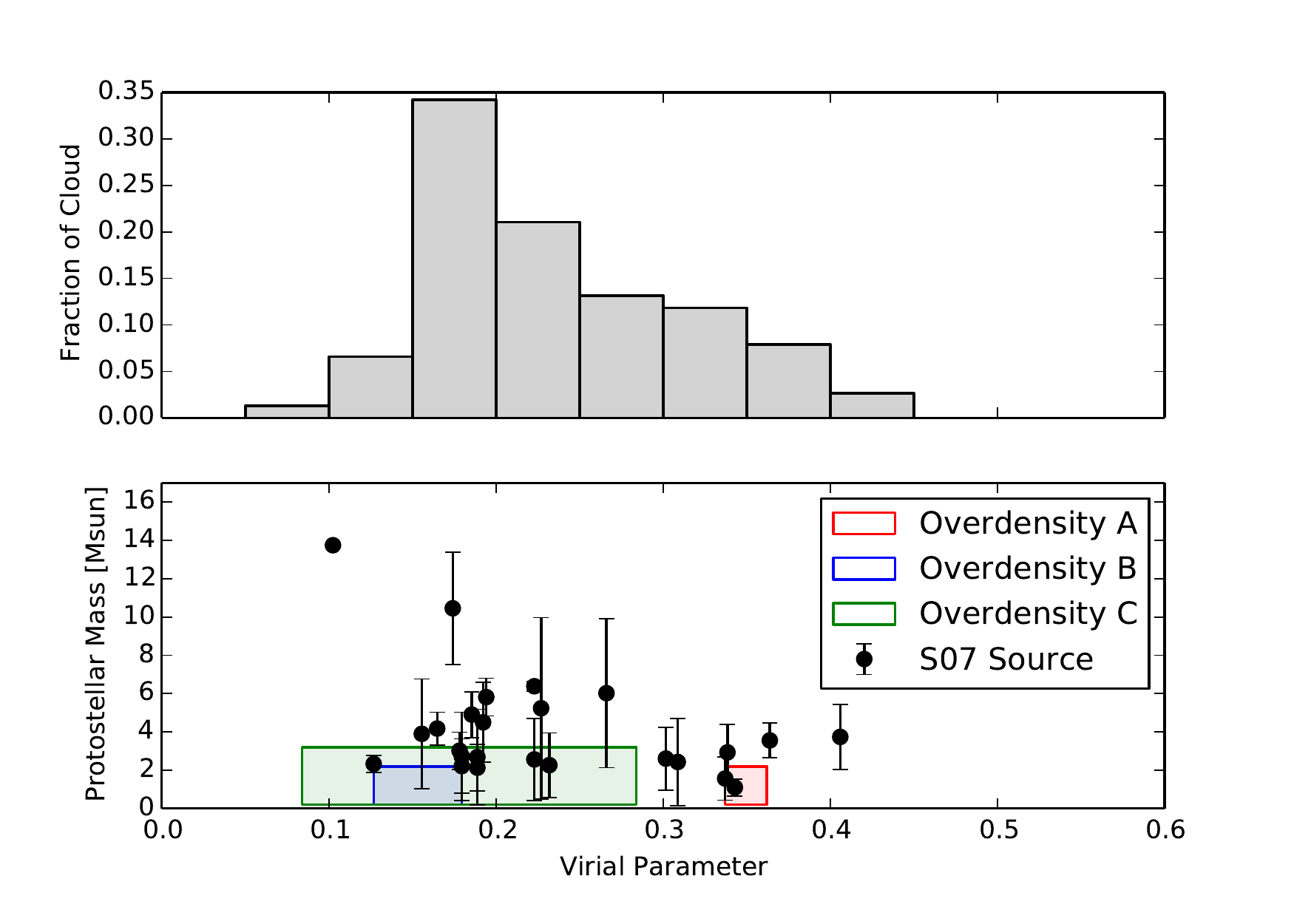}
\caption{Bottom: protostellar mass estimates and uncertainties from \citet{Shepherd:2007} vs. the virial parameter for \Gcloud. The near-infrared overdensities are shown with approximate mass ranges and the virial parameter range they cover. Note that some of the estimated uncertainties from \citet{Shepherd:2007} are smaller than the display symbol. Top: histogram of virial parameters found in \Gcloud\ of the region covered by \citet{Shepherd:2007}.}
\label{fig:mass_versus_alpha}
\end{figure}

In \autoref{fig:mass_versus_alpha} we plot the \citet{Shepherd:2007} protostellar mass estimates versus our estimate of the virial parameter, $\alpha$, in the slice of the filament where each protostar is located. The protostellar mass estimates from \citet{Shepherd:2007} are from fitting the \emph{Spitzer} fluxes to the \citet{Robitaille:2006} model grid, with the uncertainties representing the range of models in that grid that fit the fluxes; this may underestimate the uncertainty on the protostellar mass. 

For comparison, we also show the histogram of $\alpha$ for the portion of the cloud included in the \citet{Shepherd:2007} study (which excludes the southern tip of the filament; see \autoref{fig:virial_per_area}). Also shown are the near-infrared overdensities A, B, and C. Our detection of these protostars as near-infrared sources without mid-infrared detections at the level of the \citet{Shepherd:2007} study suggests that these protostars are typically of lower mass; this is true for sources of a given age (older sources would tend to have lower mid-infrared fluxes), but since the overdensities are spatially co-located with the \citet{Shepherd:2007} sources we assume they are of similar ages. Therefore, we assign a rough mass limit to the near-infrared stars in each overdensity, corresponding to the lower limit of the \citet{Shepherd:2007} masses in this region. These mass ranges and the range of virial parameters present inside the area of each overdensity are shown as the colored regions in \autoref{fig:mass_versus_alpha}.

We do not perform a quantitative analysis of these distributions because the sampling of protostellar masses is significantly incomplete and there are too few stars with well-determined masses within most of the virial parameter bins. Qualitatively, however, the high-mass protostars seem to be found in the most bound portions of the cloud ($\alpha$ $<$ 0.2). It is therefore possible that the low-mass protostars we are identifying in between the dense clumps correspond to distributed star formation in weakly bound cloud material. Follow-up observations will assess the mass function of these protostars and allow us to estimate the star formation efficiency within these regions.

\section{Conclusions}
We have detected a population of low-mass protostars in the IRDC \Gcloud. This population appears to be distributed along this filamentary dark cloud, rather than exclusively associated with the dense clumps. 

We have demonstrated the feasibility and value of detecting low-mass protostars in IRDCs using deep near-infrared images. The combination of deep near-infrared data, (lower-resolution) mid-infrared images, and interferometric (i.e., ALMA) data picks out potentially different populations of protostars and the combination allows a better view of the protostellar population than any individual method. 

We use two methods for identifying protostars in the near-infrared. Based on the column density of the cloud and our knowledge of the background stellar luminosity distribution, we assign a probability that each red ($H-K < 2$ mag) source could be a background source. By considering spatial bins of a fixed size, this allows us to identify regions where there is a significant excess of red sources over what would be expected from background sources. In addition, we identify some extended $K$-band sources where the $K$-band nebulosity arises (at least in some cases) from protostellar outflows. These two methods agree very well, with the regions of significant excess of red point sources hosting the extended $K$-band point sources. This population of protostars may explain the narrow SiO emission seen in IRDCs which appear dark and quiescent at mid-infrared wavelengths.

We have calculated the virial parameter along \Gcloud\ from a combination of submillimeter/millimeter dust continuum emission (to derive the dust temperature and mass) and a map of NH$_3$ (1,1) (to derive the turbulent line width). This virial parameter estimate allows us to distinguish regions of the filament that appear to be more or less strongly gravitationally bound. Apart from the very ends of the filament, the cloud appears to be globally gravitationally bound. There is a suggestion that the high-mass protostars are more commonly found in the regions that are most strongly bound, but additional information is required to increase the number of protostars with well-constrained masses.

The presence of distributed low-mass star formation suggests that low-mass stars may form first, and certainly may form outside of the dust clumps produced by the ``sausage'' instability in filaments. This means that some population of low-mass stars is forming under different physical conditions -- i.e., less dense and less gravitationally bound regions. Additional study is required to assess if these different conditions lead to differences in the IMF for distributed and clustered star formation. Comparison of this population with low-mass protostars in local regions of low-mass star formation could reveal differences in the star formation process as a function of environment. 

\section{Acknowledgements}

Some of the data presented herein were obtained at the W.M. Keck Observatory, which is operated as a scientific partnership among the California Institute of Technology, the University of California and the National Aeronautics and Space Administration. The Observatory was made possible by the generous financial support of the W.M. Keck Foundation. We would like to thank the observatory director, Dr. Taft Armandroff, who graciously provided his time for the acquisition of this data. The authors wish to recognize and acknowledge the very significant cultural role and reverence that the summit of Mauna Kea has always had within the indigenous Hawaiian community. We are most fortunate to have the opportunity to conduct observations from this mountain. 

A.E.G. acknowledges support from NASA grants NNX12AI55G and NNX10AD68G. This research made use of Astropy, a community-developed core Python package for astronomy \citep{astropy:2013}. This research has made use of NASA's Astrophysics Data System. This research made use of APLpy, an open-source plotting package for Python hosted at \url{http://aplpy.github.com}. The National Radio Astronomy Observatory is a facility of the National Science Foundation operated under cooperative agreement by Associated Universities, Inc. The authors wish to thank the staff at the Green Bank Telescope in particular for their assistance during observations with the KFPA. 

\bibliographystyle{apj}
\bibliography{g34.bbl}

\clearpage

\end{document}